\documentclass[aps,prd,onecolumn,showpacs,groupedaddress,nofootinbib]{revtex4-2}
\usepackage{graphicx}
\usepackage{dcolumn}
\usepackage{bm}
\usepackage{color}
\usepackage{longtable}
\usepackage{supertabular}
\usepackage[T1]{fontenc}
\usepackage{epstopdf}
\usepackage{amsmath}
\usepackage{amssymb}
\usepackage{setspace}
\usepackage{multirow}
\usepackage{booktabs}
\usepackage[section]{placeins}
\usepackage{mathrsfs}
\usepackage{hyperref}
\usepackage{adjustbox}

\makeatletter

\newcommand{\Rmnum}[1]{\expandafter\@slowromancap\romannumeral #1@} 
\newcommand{\bq}{\begin{equation}}
\newcommand{\eq}{\end{equation}}
\newcommand{\bqn}{\begin{eqnarray}}
\newcommand{\eqn}{\end{eqnarray}}
\newcommand{\nb}{\nonumber}

\makeatother

\begin{document}
\title{On the universality of instability in the fundamental quasinormal modes of black holes}

\author{Wei-Liang Qian\textsuperscript{2,1,3}}\email[E-mail: ]{wlqian@usp.br (corresponding author)}
\author{Guan-Ru Li\textsuperscript{3}}\email[E-mail: ]{guanru.li@unesp.br}
\author{Ramin G. Daghigh\textsuperscript{4}}\email[E-mail: ]{ramin.daghigh@metrostate.edu}
\author{Stefan Randow\textsuperscript{5}}\email[E-mail: ]{stefan.randow@my.metrostate.edu}
\author{Rui-Hong Yue\textsuperscript{1}}\email[E-mail: ]{rhyue@yzu.edu.cn}

\affiliation{$^{1}$ Center for Gravitation and Cosmology, College of Physical Science and Technology, Yangzhou University, Yangzhou 225009, China}
\affiliation{$^{2}$ Escola de Engenharia de Lorena, Universidade de S\~ao Paulo, 12602-810, Lorena, SP, Brazil}
\affiliation{$^{3}$ Faculdade de Engenharia de Guaratinguet\'a, Universidade Estadual Paulista, 12516-410, Guaratinguet\'a, SP, Brazil}
\affiliation{$^{4}$ Natural Sciences Department, Metropolitan State University, Saint Paul, Minnesota, 55106, USA}
\affiliation{$^{5}$ Mathematics and Statistics Department, Metropolitan State University, Saint Paul, Minnesota, 55106, USA}

\begin{abstract}
We elaborate on a criterion for the emergence of instability in the fundamental mode recently observed by Cheung {\it et al.}, as a universal phenomenon in the context of black hole perturbations.
Such instability is characterized by an exponential spiral, deviating from the quasinormal frequencies due to an insignificant perturbation moving away from the compact object.
Our analysis begins with a specific case involving a truncated P\"oschl-Teller potential for which we derive an explicit form of the criterion.
Notably, it is shown analytically, contrary to other cases studied in the literature, that the fundamental mode is stable.
These derivations are then generalized to a broader context, embracing two underlying mathematical formalisms.
Specifically, the spiral is attributed to either the poles in the black hole's reflection amplitude or the zeros in the transmission amplitude.
Additionally, we revisit and then generalize a toy model in which perturbations to the effective potential are disjointed, demonstrating that such a configuration invariably leads to instability in the fundamental mode, and the resulting outward spiral always occurs in the counter-clockwise direction.
The derived criterion is not restricted to the fundamental mode but is generally relevant for the first few low-lying modes.
We demonstrate numerically that the spiral's period and the frequency's relative deviation agree well with our analytical estimations.
\end{abstract}

\date{Dec. 11th, 2024}

\maketitle

\newpage
\section{Introduction}\label{sec1}

Black holes are one of the most captivating concepts in theoretical physics. They exemplify gravity's properties at their extremity. 
The successful detection of gravitational waves emanating from the binary mergers achieved by the ground-based LIGO and Virgo collaboration~\cite{agr-LIGO-01, agr-LIGO-02, agr-LIGO-03, agr-LIGO-04} marked the inauguration of a new epoch for observational astrophysics. 
This significant milestone has further inspired the ongoing spaceborne projects, such as LISA~\cite{agr-LISA-01}, TianQin~\cite{agr-TianQin-01, agr-TianQin-Taiji-review-01}, and Taiji~\cite{agr-Taiji-01}, fostering the speculation that direct observation of ringdown waveforms is plausible~\cite{agr-TianQin-05}.

These particular waveforms predominantly consist of quasinormal modes (QNMs)~\cite{agr-qnm-review-02, agr-qnm-review-03, agr-qnm-review-06}, whose study has sparked considerable interest. 
It is well-known, as noted by Leaver~\cite{agr-qnm-21, agr-qnm-29}, that QNMs are linked with the poles of Green's function related to the master wave equation, whereas the late-time behavior is associated with the branch cut, typically demonstrating a decay that follows an inverse power in time~\cite{agr-qnm-tail-01}. 
The unique characteristic of QNMs, determined solely by the spacetime properties surrounding the black hole, provides clear and distinct information about the spacetime geometry near the event horizon.

The behavior and properties of QNMs are intrinsically linked with the concept of spectral instability. 
The latter was pioneered by Nollert and Price~\cite{agr-qnm-35, agr-qnm-36} and Aguirregabiria and Vishveshwara~\cite{agr-qnm-27, agr-qnm-30}.
It was shown that even minor perturbations, such as step functions, can qualitatively influence the higher overtones in the QNM spectrum, demonstrating an unexpected instability of the QNM spectrum against so-called ``ultraviolet'' perturbations. 
This finding challenges the conventional assumption that a reasonable approximation of the effective potential ensures minimal deviation in the resulting QNMs. 
Further investigations~\cite{agr-qnm-50, agr-qnm-lq-03, agr-qnm-echoes-20} have suggested that even in the presence of discontinuities of a more moderate nature, the asymptotic behavior of the QNM spectrum could be non-perturbatively modified. 
Specifically, high-overtone modes might shift along the real frequency axis rather than ascending the imaginary frequency axis observed for most black hole metrics~\cite{agr-qnm-continued-fraction-12, agr-qnm-continued-fraction-23}. 
This phenomenon persists regardless of the discontinuity's distance from the horizon or its magnitude.
Expanding on this concept, Jaramillo {\it et al.}~\cite{agr-qnm-instability-07, agr-qnm-instability-13} explored the implications of spectral stability by analyzing the effects of randomized perturbations to the metric in terms of the notion of pseudospectrum in the context of black hole perturbation theory.
Their analyses revealed that the boundary of the pseudospectrum moves closer to the real frequency axis, thus reinforcing the notion of a universal instability of high-overtone modes triggered by ultraviolet perturbations.

The study of spectral instability and its ramifications has significant observational implications, especially for black hole spectroscopy~\cite{agr-bh-spectroscopy-05, agr-bh-spectroscopy-06, agr-bh-spectroscopy-15, agr-bh-spectroscopy-18, agr-bh-spectroscopy-20, agr-bh-spectroscopy-36}. 
In real-world astrophysical contexts, gravitational radiation sources such as black holes or neutron stars are seldom isolated; they are typically submerged and interacting with surrounding matter. 
This leads to deviations from the ideal symmetric metric, thereby causing the emitted gravitational waves of the underlying QNMs to differ substantially from those predicted for a pristine, isolated, compact object. 
Such phenomena have motivated the investigation into ``dirty'' black holes, as explored by several authors~\cite{agr-bh-thermodynamics-12, agr-qnm-33, agr-qnm-34, agr-qnm-54}, and opened new avenues in the study of black hole perturbation theory.
Moreover, the asymptotic modes that align almost parallel to the real axis are closely related to the intriguing concept of echoes, a late-stage ringing waveform first proposed by Cardoso {\it et al.}~\cite{agr-qnm-echoes-01}. 
As potential observables, echoes might help distinguish different but otherwise similar gravitational systems via their distinct properties near the horizon. 
This idea has spurred many studies into echoes across various systems, encompassing exotic compact objects such as gravastars~\cite{agr-eco-gravastar-02, agr-eco-gravastar-03}, boson stars~\cite{agr-eco-gravastar-07}, and wormholes~\cite{agr-wormhole-01, agr-wormhole-02, agr-wormhole-10, agr-wormhole-11}. 
Like the late-time tail, echoes are also attributed to the analytic properties of Green's function, as analyzed by Mark {\it et al.}~\cite{agr-qnm-echoes-15}. 
In studies of Damour-Solodukhin type wormholes\cite{agr-wormhole-12}, Bueno {\it et al.}~\cite{agr-qnm-echoes-16} have investigated echoes by explicitly solving for specific frequencies at which the transition matrix becomes singular, providing further insights into the complex interplay of spacetime geometry and QNMs.

The related topic of the spectral instability, echoes, and causality has been explored extensively in recent years by many authors~\cite{agr-qnm-instability-08, agr-qnm-instability-13, agr-qnm-instability-14, agr-qnm-instability-15, agr-qnm-instability-16, agr-qnm-instability-18, agr-qnm-instability-19, agr-qnm-instability-26, agr-qnm-echoes-22, agr-qnm-echoes-29, agr-qnm-echoes-30, agr-qnm-instability-23, agr-qnm-instability-29, agr-qnm-instability-32, agr-qnm-instability-33, agr-qnm-instability-43, agr-qnm-echoes-35}.
Notably, Cheung {\it et al.}~\cite{agr-qnm-instability-15} pointed out that even the fundamental mode can be destabilized under generic perturbations.
By introducing a small perturbation to the Regge-Wheeler effective potential, it was shown~\cite{agr-qnm-instability-15} that the fundamental mode undergoes an outward spiral while the deviation's magnitude increases.
Subsequently, its position is overtaken by a new mode distinct from the first few low-lying modes. 
The results are ascertained using different shapes for the perturbative bump and observed in a toy model constituted by two disjointed rectangular potential barriers~\cite{agr-qnm-instability-15, agr-qnm-instability-32}.
Such an observation undermines the understanding that spectral instability might not significantly impact black hole spectroscopy as the fundamental mode is not subjected to spectral instability, leading to substantial observational implications.
The physical significance of such instability has been further scrutinized, inclusively regarding a different scale introduced to the problem via the metric perturbations, in more recent studies~\cite{agr-qnm-instability-47, agr-qnm-instability-56, agr-qnm-instability-59}.

While inspiring, existing studies~\cite{agr-qnm-instability-15} primarily reside on numerical calculations with unprecedented precision.
The present study is motivated to explore this topic further, along with the existing efforts~\cite{agr-qnm-instability-57, agr-qnm-instability-58} from the analytic perspective.
Our analysis consists of the following key ingredients:
\begin{itemize}
    \item We consider a small perturbation to the original black hole metric\footnote{An appropriate approach would be to derive the modification to the effective potential by introducing some specific perturbations to the metric.
    Instead, in the present study, similar to~\cite{agr-qnm-instability-15}, the perturbations are directly assigned to the effective potential whose form is given by hand.
    In this regard, the current approach is more of a toy model that aims to simplify the underlying mathematics, serving as a preliminary attempt to capture the essence of the underlying physics.}.
    \item The discussions primarily focus on the fundamental mode.
    As it turns out, however, our arguments are also pertinent to the first few low-lying modes.
    \item For mathematical feasibility, we assume that the deviation of the fundamental mode, triggered by the perturbation, is not large\footnote{As shown below in Sec.~\ref{sec2}, the deviation may increase exponentially, quickly invalidating this assumption. As a result, in theory, such analysis is limited to assessing linear instability. Nevertheless, numerical calculations suggest that the deviation of the QNM can be substantial before the analytical estimation ceases to be valid.}.
    \item We assess the evolution of QNMs as the perturbation's location moves away from the black hole.
    Based on the above assumption, the spiral, if any, takes place in the vicinity of the original location of QNMs on the complex frequency plane.
\end{itemize}
To this end, we primarily elaborate on two classes of black hole metrics associated with different underlying mathematical formalisms.
To be specific, the QNMs are either due to the poles of the reflection amplitude or they are attributed to the zeros of the transmission amplitude.
We start with a simplified scenario that involves a small discontinuity, such as a step, introduced into the original black hole's effective potential away from the horizon.
Explicit calculations are given for an analytically attainable model, a truncated P\"oschl-Teller potential.
Subsequently, instead of a discontinuity, we consider that the perturbation could be of an arbitrary finite shape 
This is initiated by exploring a group of models where the original black hole's metric and that of the perturbations are defined on separated compact domains.
By putting all the pieces together, we elaborate on the criteria for the emergence of instability in the low-lying modes in a general context.

Here, we consider the theoretical setup in which the study of black hole perturbations can be simplified by exploring the radial part of the master equation~\cite{agr-qnm-review-03,agr-qnm-review-06},
\begin{eqnarray}
\frac{\partial^2}{\partial t^2}\Psi(t, x)+\left(-\frac{\partial^2}{\partial x^2}+V_\mathrm{eff}\right)\Psi(t, x)=0 ,
\label{master_eq_ns}
\end{eqnarray}
where the radial coordinate is chosen to be the tortoise coordinate $x$, the effective potential $V_\mathrm{eff}$ is determined by the given spacetime metric, spin ${\bar{s}}$, and angular momentum $\ell$ of the perturbation.
For instance, the Regge-Wheeler potential for a Schwarzschild black hole with mass $M$ reads
\bqn
V_\mathrm{RW}=f\left[\frac{\ell(\ell+1)}{r^2}+(1-{\bar{s}}^2)\frac{r_h}{r^3}\right],
\label{V_RW}
\eqn
where 
\bqn
f=1-r_h/r ,
\label{f_master}
\eqn
and the horizon $r_h=2M$.

The quasinormal frequencies can be obtained by evaluating the zeros of the Wronskian
\begin{eqnarray}
W(\omega)\equiv W(g,f)=g(\omega,x)f'(\omega, x)-f(\omega,x)g'(\omega,x) ,
\label{pt_Wronskian}
\end{eqnarray}
where $'\equiv d/dx$, and $f$ and $g$ are the solutions of the corresponding homogenous equation of Eq.~\eqref{master_eq_ns} in frequency domain~\cite{agr-qnm-review-02},
\begin{eqnarray}
\left[-\omega^2-\frac{d^2}{dx^2}+V_\mathrm{eff}\right]\widetilde{\Psi}(\omega, x)=0 ,
\label{pt_homo_eq}
\end{eqnarray}
with appropriate boundary conditions, namely,
\begin{eqnarray}
\begin{array}{cc}
f(\omega, x)\sim e^{-i\omega x}    &  x\to -\infty  \cr\\
g(\omega, x)\sim e^{i\omega x}     &  x\to +\infty  
\end{array} ,
\label{pt_boundary}
\end{eqnarray}
As discussed above, the present paper involves the choice of physically relevant effective potential $V_\mathrm{eff}$ and their implication for the instability of the low-lying QNMs.

The remainder of the paper is organized as follows.
In the following two sections, we elaborate on the properties and criteria for the emergence of instability in the low-lying modes.
In Sec.~\ref{sec2}, we start our analysis with a truncated P\"oschl-Teller potential, from which we derive the sufficient condition for the instability.
These derivations are then generalized to a broader context where the black hole QNMs are attributed either to the poles of the reflection amplitude or the zeros of the transmission amplitude, while the perturbation is simplified by a truncation in the effective potential.
In Sec.~\ref{sec3}, we further extend our discussion by considering the case where the perturbation in the effective potential possesses an arbitrary finite profile.
To this end, we revisit a toy model in which perturbations to the effective potential are disjointed, demonstrating that such a configuration always leads to instability in the fundamental mode.
Subsequently, we elaborate on the general scenario in which an arbitrary small perturbation is introduced to a continuous black hole metric.
Numerical calculations supporting our theoretical findings are presented in Sec.~\ref{sec4}.
The last section includes further discussions and concluding remarks.

\section{Instability in the truncated effective potentials}\label{sec2}

This section elaborates on a specific class of perturbations corresponding to an infinitely thin mass shell wrapped around a central black hole.
Our derivation will focus on the case where the presence of the matter gives rise to a discontinuity in the metric.
It is worth noting that, to our knowledge, most examples reported the instability in the fundamental mode~\cite{agr-qnm-instability-15, agr-qnm-instability-56, agr-qnm-instability-58}, which involves some discontinuity in the metric perturbation.
Without loss of generality, we elaborate on a scenario that significantly simplifies the mathematical formulation.
Specifically, one will consider the case where the effective potential is truncated at a given position $x_\mathrm{cut}$.
Moreover, as discussed below, its generalization to more realistic cases is straightforward.

\subsection{The truncated P\"oschl-Teller potential}\label{sec2.1}

We will first consider an explicit form of the P\"oschl-Teller potential, for which the derivation can be primarily performed analytically.
Such an effective potential has been extensively discussed in the literature~\cite{agr-qnm-Poschl-Teller-01, agr-qnm-Poschl-Teller-02, agr-qnm-Poschl-Teller-03, agr-qnm-Poschl-Teller-04, agr-qnm-Poschl-Teller-05, agr-qnm-lq-03} due to its resemblance to the Regge-Wheeler potential.
To introduce the perturbation, we will truncate the potential at a large radial coordinate $x=x_\mathrm{cut}$, far away from the potential's maximum.
Specifically, the effective potential reads
\bqn
\widetilde{V}_\mathrm{PT}=
\left\{\begin{array}{cc}
{V}_\mathrm{PT}(x)     &  x\le x_\mathrm{cut}  \cr\\
0    &  x> x_\mathrm{cut}  
\end{array}\right. ,
\label{V_mpt}
\eqn
which is constructed by introducing a ``cut'' at $x_\mathrm{cut} $ in the P\"oschl-Teller potential,
\begin{eqnarray}
{V}_\mathrm{PT}=\frac{V_m}{\cosh ^2(\kappa x)} ,\label{V_PT}
\end{eqnarray}
where $V_m$ and $\kappa$ are constants.

To proceed, we first evaluate $f$ and $g$ and then explore the zeros of the Wronskian Eq.~\eqref{pt_Wronskian}.
Based on the well-known method~\cite{agr-qnm-Poschl-Teller-01,agr-qnm-Poschl-Teller-02}, the QNMs for the original (inverse) P\"oschl-Teller potential can be directly obtained from the eigenvalues of the corresponding bound-state problem where $\omega$ is real.
Specifically, for the modified potential Eq.~\eqref{V_mpt}, the wave function $f(\omega,x)$ of the potential $\widetilde{V}_\mathrm{PT}$ for $x<x_\mathrm{cut}$ can be obtained from its bound state counterpart by using the transformation 
\bqn
\left\{\begin{array}{c}
x\to -ix  \cr\\
\kappa \to i\kappa  
\end{array}\right. .
\label{trans_PT}
\eqn
Generally, after the above transformation, the waveform does not vanish at spatial infinity on the imaginary axis.
Nonetheless, in order to satisfy the in-going boundary condition governed by the first line of Eq.~\eqref{pt_boundary}, one must choose a proper combination of the analytic forms of the corresponding bound state solutions with well-defined parities~\cite{book-quantum-mechanics-Flugge}.
The resulting form of $f(\omega,x)$ can be written as
\bqn
f(\omega,x) = A u_e + B u_o . 
\label{f_mPT}
\eqn
We delegate the specific forms of $u_e, u_o$, and the coefficients $A, B$ to the Appendix.

Owing to the truncation, the wave function $g(\omega, x)$ for $x>x_\mathrm{cut}$ satisfying the boundary condition Eq.~\eqref{pt_boundary} has the form
\bqn
g(\omega,x) = e^{i\omega x} . 
\label{g_mPT}
\eqn
By using Eqs.~\eqref{f_mPT} and~\eqref{g_mPT}, the quasinormal frequencies can be obtained by analyzing the roots of the Wronskian Eq.~\eqref{pt_Wronskian}.

Given a large but finite $x_\mathrm{cut}$, some approximation has to be adopted to proceed further.
For instance, semi-analytic results for the asymptotical quasinormal frequencies are obtained in~\cite{agr-qnm-Poschl-Teller-03, agr-qnm-Poschl-Teller-04, agr-qnm-lq-03}.
Since asymptotical QNMs are involved, either the real or the imaginary part of the quasinormal frequency is large. 
As a result, one employs Bailey's theorem~\cite{agr-qnm-Poschl-Teller-05} to write down the waveform in terms of the Gamma function and applies the reflection formula to transform the argument into the appropriate region so that the asymptotic expansion of the Gamma function can be utilized.
For the fundamental mode, however, the corresponding argument of the Gamma function is not large.
Nonetheless, as elaborated below, we can take advantage of the properties that the frequency in question is in the immediate vicinity of the black hole's lowest-lying modes, namely, the poles of $C/D$, which possess a small imaginary part.

Before proceeding further, we note if one takes the limit $x_\mathrm{cut}\to +\infty$ in the wave functions $u_e, u_o$ and then substitutes Eq.~\eqref{f_mPT} into the Wronskian,
one finds
\bqn
f(\omega,x) \to C e^{i\omega x} + D e^{-i\omega x} \ \ \ \mathrm{for}\ x\to+\infty, 
\label{f_mPT_CD}
\eqn
where $C$ and $D$ can be interpreted as the reflection and transmission amplitudes of the black hole, and their specific forms are given by Eq.~\eqref{CD_ueuo}.
One observes that the Wronskian vanishes at the poles of $C/D$.
By using Eq.~\eqref{CD_ueuo}, it corresponds to any factor of the product $\Gamma(b)\Gamma\left(b+\frac12\right)\Gamma\left(1-a\right)\Gamma\left(\frac12-a\right)$ to diverge.
Subsequently, the quasinormal frequencies (with positive real parts) are found to be
\bqn
\omega_n = \omega_R + i \omega_I ,  \label{qnm_omega} 
\eqn
where $n$ is a non-negative integer and
\bqn
\omega_R=\omega^{\mathrm{PT}}_R &=& \sqrt{V_m-\frac{\kappa^2}{4}},\nonumber \\
\omega_I=\omega^{\mathrm{PT}}_I &=& -\left(n+\frac12\right)\kappa .
\label{qnm_PT}
\eqn
This result is nothing but the original P\"oschl-Teller potential~\cite{agr-qnm-Poschl-Teller-01,agr-qnm-Poschl-Teller-02} where the spectrum climbs down along the imaginary axis.
Therefore, although $x_\mathrm{cut}$ is large, one must postpone taking the limit $x_\mathrm{cut}\to \infty$ in the derivations. 

In this regard, owing to the ``cut'' implemented at a finite location, the Wronskian should not be evaluated at asymptotic spatial infinity but at $x=x_\mathrm{cut}$. 
We aim to estimate how such a modification gives rise to a tiny deviation from the poles of the original black hole, namely, the first line of Eq.~\eqref{CD_ueuo}.
Specifically, at the original quasinormal frequency, the Wronskian between the wave functions $f(\omega, x)$ and $g(\omega, x)$ given by Eqs.~\eqref{f_mPT} and~\eqref{g_mPT}, which can be conveniently evaluated at $x=x_\mathrm{cut}$, does not vanish.
The deviation of the QNMs resides in how to estimate that in the wave function $f(\omega, x)$.
Following the above discussion, if one approximates $f(\omega, x)$ to the first order and utilizes Eq.~\eqref{f_mPT_CD}, we do not find any deviation in the quasinormal frequency.
Fortunately, desirable results can be obtained if one proceeds and expands the wave function at $x\to +\infty$ to the first order. 
Instead of Eq.~\eqref{f_mPT_CD}, we have
\bqn
f(\omega,x) \to \widetilde{C} e^{i\omega x} + \widetilde{D} e^{-i\omega x} \ \ \ \mathrm{for}\ x\to+\infty, 
\label{f_mPT_CD_tilde}
\eqn
where
\bqn
\widetilde{C} &\equiv& C+\Delta C=C + \left(\Delta \widetilde{C}_1 + \Delta \widetilde{C}_2 \right)e^{-2\kappa x}, \nonumber \\
\widetilde{D} &\equiv& D+\Delta D=D + \Delta \widetilde{D} e^{-2\kappa x} ,\label{CD_tilde_ueuo}
\eqn
and the specific forms of the coefficients $\Delta \widetilde{C}_1$, $\Delta \widetilde{C}_2$, and $\Delta \widetilde{D}$ are given in Eqs.~\eqref{CD_tilde_delta_ueuo}.
Owing to the exponential factor $e^{-2\kappa x}$, the corrections manifestly vanish at spatial infinity 
\bqn
\lim\limits_{x\to\infty}\Delta C=\lim\limits_{x\to\infty}\Delta D = 0 .\label{DeltaCDasymp}
\eqn

The resultant Wronskian reads
\bqn
W(\omega)= \Delta W(\omega) = \Delta W_C(\omega) + \Delta W_D(\omega). \label{Wronskian_PT_dev}
\eqn
where 
\bqn
\Delta W_C(\omega) \doteq W(g, \Delta C e^{i\omega x}) = -2\kappa (\Delta\widetilde{C}_1+\Delta\widetilde{C}_2) e^{-2\kappa x_\mathrm{cut}}e^{2i\omega x_\mathrm{cut}}
\label{Wronskian_PT_dev_C}
\eqn
and 
\bqn
\Delta W_D(\omega) \doteq W(g, D e^{-i\omega x}) = -2i\omega D
\label{Wronskian_PT_dev_D}
\eqn

Equating Eq.~\eqref{Wronskian_PT_dev} to zero at the truncation point $x=x_\mathrm{cut}$ gives an algebraic equation for the quasinormal frequencies $\omega$.
In the vicinity of a given low-lying mode, $\omega_n$, the deviation $\omega-\omega_n$ is a small quantity.
It will significantly impact the term related to $\Delta {C}$ since $\omega_n$ is a pole.
Subsequently, the deviation is also sizable.
To properly estimate the contribution, we divide both terms on the r.h.s. of Eq.~\eqref{Wronskian_PT_dev} by $C$.
It can be shown that $\omega_n$ is a simple pole of $C(\omega)$, namely, $\mathrm{Res}\left(\Gamma, -n\right)=(-1)^n/n!$
Subsequently, we have
\bqn
\lim\limits_{\omega\to\omega_n}\frac{1}{(\omega-\omega_c)}\frac{D}{C} = \frac{D(\omega_n)}{\mathrm{Res}\left(C,\omega_n\right)} 
=\left\{\begin{array}{cc}
\frac{i(-1)^{(n+1)} n!\Gamma\left(-\frac{i\omega_n}{\kappa} \right)}{\kappa\Gamma\left(\frac{i\omega_n}{\kappa} \right)}\frac{\Gamma\left(\frac{\lambda}{2}+i\frac{\omega_n}{2\kappa}\right)\Gamma\left(\frac{1-\lambda}{2}+i\frac{\omega_n}{2\kappa}\right)}{\Gamma\left(\frac{1-\lambda}{2}-i\frac{\omega_n}{2\kappa}\right)} & \ \ \mathrm{for}\ n=0,2,\cdots  \cr\\
\frac{i(-1)^{(n+1)} n!\Gamma\left(-\frac{i\omega_n}{\kappa} \right)}{\kappa\Gamma\left(\frac{i\omega_n}{\kappa} \right)}\frac{\Gamma\left(\frac{\lambda+1}{2}+i\frac{\omega_n}{2\kappa}\right)\Gamma\left(\frac{2-\lambda}{2}+i\frac{\omega_n}{2\kappa}\right)}{\Gamma\left(\frac{2-\lambda}{2}-i\frac{\omega_n}{2\kappa}\right)} & \ \ \mathrm{for}\ n=1,3,\cdots
\end{array}\right. ,\label{DResC}
\eqn
where $\mathrm{Res}\left(C,\omega_n\right)$ is the residual of $C(\omega)$ at the singularity $\omega=\omega_n$ given by Eq.~\eqref{ResComegan}.
Also, $\Delta C/C$ is manifestly finite at $\omega\to \omega_c$ by using the explicit forms of Eqs.~\eqref{CD_ueuo} and~\eqref{CD_tilde_delta_ueuo}, 
\bqn
\lim\limits_{\omega\to\omega_n}\left(\frac{\Delta\widetilde{C}_1+\Delta\widetilde{C}_2}{C}\right)
= \left\{
\begin{array}{cc}
\frac{\Delta\widetilde{C}_1}{C} & \ \ \mathrm{for}\ n=0,2,\cdots  \cr\\
\frac{\Delta\widetilde{C}_2}{C} & \ \ \mathrm{for}\ n=1,3,\cdots
\end{array}\right\} 
= -\frac{\lambda(\lambda-1)}{\left(1-i\frac{\omega_n}{\kappa}\right)}, \label{DeltaTCC}
\eqn
where one obtains identical results after the poles in the numerator and denominator cancel out.
As a result, both $D/C$ and $\Delta C/C$ are now well-defined, and the vanishing condition of the Wronskian can be reformulated in terms of the deviation from the QNM in question
\bqn
\delta\omega_n^{\mathrm{pole}}\equiv \omega-\omega_n = \mathcal{J}_n e^{(2i\omega_n-2\kappa) x_\mathrm{cut}}
=\mathcal{J}_n e^{\left(2n-1\right)\kappa x_\mathrm{cut}} e^{i 2\sqrt{V_m-\frac{\kappa^2}{4}} x_\mathrm{cut}} ,\label{fundamentalInsPT}
\eqn
where
\bqn
\mathcal{J}_n = \frac{i\kappa}{\omega_n} \frac{\mathrm{Res}\left(C,\omega_n\right)}{D(\omega_n)}\lim\limits_{\omega\to\omega_n}\left(\frac{\Delta\widetilde{C}_1+\Delta\widetilde{C}_2}{C}\right) ,\label{JformPole1}
\eqn
as given by Eqs.~\eqref{DResC} and~\eqref{DeltaTCC}, which can be treated primarily as a constant in the immediate vicinity $\omega\sim \omega_n$.

Now, the last equality of Eq.~\eqref{fundamentalInsPT} can be readily used to demonstrate the criterion for the possible instability of the low-lying modes.
As long as $\mathcal{J}_n$ is moderate, the exponential factor $e^{\left(2n-1\right)\kappa x_\mathrm{cut}}$ primarily governs the magnitude of the deviation $\delta\omega_n$, while the factor $e^{i \sqrt{V_m-\frac{\kappa^2}{4}} x_\mathrm{cut}}$ control its phase.
As the phase $\sqrt{V_m-\frac{\kappa^2}{4}} x_\mathrm{cut}$ increase linearly with $x_\mathrm{cut}$, the magnitude of the deviation increases or decreases exponentially, giving rise to the spiral.
Since $\kappa>0$, as long as $n> \frac12$, the spiral will occur and grow exponentially as $x_\mathrm{cut}$ increases.
For $\omega_R >0$, the spiral rotates in the counter-clockwise direction.
It is straightforward to see that the phenomenon becomes more pronounced, mainly when $\omega_I$ is relatively small compared to $\omega_R$.
In this regard, the instability is more relevant for the low-lying modes, except that the fundamental mode $(n=0)$ will remain stable against the perturbation. 
On the other hand, for high overtones $n\gg 1$, the deviation would be too large so that the assumptions used for the limits in deriving Eqs.~\eqref{DResC} and~\eqref{DeltaTCC} will break down.

\subsection{The generalization to an arbitrary truncated effective potential}\label{sec2.2}

It is inviting to borrow the spirit of the derivation for P\"oschl-Teller potential to more general cases.
First of all, the asymptotical forms given by Eqs.~\eqref{pt_boundary} and~\eqref{f_mPT_CD}, namely, 
\begin{eqnarray}
\begin{array}{cc}
f(\omega, x)\sim e^{-i\omega x}    &  x\to -\infty  ,
\end{array} 
\end{eqnarray}
and
\bqn
f(\omega,x) \to C e^{i\omega x} + D e^{-i\omega x} \ \ \ \mathrm{for}\ x\to+\infty, \nb
\eqn
remain unchanged for any effective potential that vanishes at spatial infinity
\bqn
\lim\limits_{x\to\infty}V_\mathrm{eff} = 0 ,
\eqn
where the specific forms of the coefficients are determined by the specific form of the effective potential.

As the effective potential is truncated at $x=x_\mathrm{cut}$, the out-going wave $g(\omega, x)$ is given by Eq.~\eqref{g_mPT} for $x>x_\mathrm{cut}$
\bqn
g(\omega,x) = e^{i\omega x} . \nb
\eqn

The Wronskian can be evaluated at $x=x_\mathrm{cut}$, whose zeros, by definition, are essentially governed by the poles of $C/D$.
This leaves us with two possibilities: the QNMs are associated with either the poles of $C$ or the zeros of $D$.
For the truncated P\"oschl-Teller potential discussed in the last subsection, as the Gamma function contains no zeros, neither does the reflection amplitude $D$.
In this case, the QNMs are attributed to the poles in the reflection amplitude $C$. 
In what follows, we first generalize the specific case of truncated P\"oschl-Teller potential given in the last subsection, aiming at a generic proof for the first class of models.
Subsequently, we elaborate on the second class of models.

For the first class of black hole metrics, we will primarily borrow the strategy employed in Sec.~\ref{sec2.1}.
As discussed, the limit $x_\mathrm{cut} \to +\infty$ should be postponed in the derivation.
To assess the deviation due to the ``cut'' at $x_\mathrm{cut}$, one again resort to the first-order form Eq.~\eqref{f_mPT_CD_tilde}
\bqn
f(\omega,x) \to \widetilde{C} e^{i\omega x} + \widetilde{D} e^{-i\omega x}\equiv (C+\Delta C) e^{i\omega x} + D e^{-i\omega x} \ \ \ \mathrm{for}\ x\to+\infty . \nb
\eqn
As the zeroth order, $C$ is not a function of the radial coordinate.
When compared $\Delta C$ and $C$, we have the limit
\bqn
\lim\limits_{x\to \infty}\frac{\Delta C}{C} = 0 .
\eqn
This implies $\Delta C$ explicitly depends on $x$, and the function is suppressed as $x\to \infty$.
For example, in the case of P\"oschl-Teller potential, such a suppression is materialized by the factor $e^{-2\kappa x}$ (c.f. Eqs.~\eqref{CD_tilde_ueuo} and~\eqref{CD_tilde_delta_ueuo}).

As a result, the modification to the Wronskian consists of two dominant contributions.
The first one is owing to the above-mentioned radial-coordinate dependence.
Specifically, we have
\bqn
\Delta W_C \doteq W\left(g,\Delta Ce^{i\omega x}\right) = \left.\frac{d\left(\Delta C\right)}{dx}\right|_{x=x_\mathrm{cut}} e^{2i\omega x_\mathrm{cut}} .
\eqn
It is noted even though $\frac{d\left(\Delta C\right)}{dx}\ne 0$, since $\Delta C$ is a correction of the order $1/x$, it has the asymptotical form
\bqn
\lim\limits_{x\to \infty}\frac{d}{dx}\Delta C = 0 ,\label{assDeltaC}
\eqn
similar to Eq.~\eqref{DeltaCDasymp}.

The second contribution is owing to the coefficient $D$ 
\bqn
\Delta W_D \doteq W\left(g,De^{-i\omega x}\right) = -2i\omega D .
\eqn
The contribution from $D$ to the Wronskian was entirely irrelevant when $C$ attains a pole, but it plays a role when the frequency slightly deviates from $\omega_n$.
In particular, the Wronskian vanishes identically when the above two contributions precisely cancel out, giving rise to an equation that determines the deviation $\delta\omega_n\equiv \omega-\omega_n$.

Similar to the approach employed in Sec.~\ref{sec2.1}, one divides both terms by $C$.
On the one hand, the term $D/C$ is dominated by the pole associated with the QNM.
If it is a simple pole, we have Eq.~\eqref{DResC}
\bqn
\lim\limits_{\omega\to\omega_n}\frac{1}{(\omega-\omega_c)}\frac{D}{C} = \frac{D(\omega_n)}{\mathrm{Res}\left(C,\omega_n\right)} ,\nb
\eqn
where the specific forms of the residual and coefficient depend on a particular black hole metric.
On the other hand, the limit of the ratio $\Delta C/C$ is a well-defined quantity as long as the poles in $\Delta C$ and $C$ cancel out identically.
It seems to be a coincidence for the particular example of the P\"oschl-Teller potential.
However, such an apparent coincidence can partly be understood in the sense that $\Delta C$ and $C$ are quantities of different orders in a series expansion in terms of $1/x$ but not in frequency.
Nonetheless, it is an educated assumption that the following limit exists
\bqn
\frac{d}{dx}\left(\lim\limits_{\omega\to\omega_n}\frac{\Delta C}{C}\right) .
\eqn

By putting all the pieces together, one arrives at the following expression for the frequency deviation, which strongly resembles Eq.~\eqref{fundamentalInsPT}
\bqn
\delta\omega_n^{\mathrm{pole}} =\mathcal{J}_n e^{\left(-2\omega_I+\Re\mathcal{K}_n\right) x_\mathrm{cut}} e^{i\left( 2\omega_R +\Im\mathcal{K}_n\right)x_\mathrm{cut}} ,\label{fundamentalInsGenPole}
\eqn
where
\bqn
\mathcal{J}_n &=& \frac{i}{2\omega_n}\frac{\mathrm{Res}\left(C,\omega_n\right)}{D(\omega_n)} ,\nb\\
\mathcal{K}_n &=& \frac{1}{x_\mathrm{cut}}\ln\left.\left[\frac{d}{dx}\lim\limits_{\omega\to\omega_n}\frac{\Delta C}{C}\right]\right|_{x=x_\mathrm{cut}} .
\eqn
Most discussions regarding the stability of low-lying modes established in the preceding section can be readily adapted to the general case without much modification.
A given black hole metric's characteristic resides in the term $\mathcal{K}_n$.
The logarithmic operation involved in its definition is due to the significantly suppressed correction $\Delta C$ at large spatial coordinates (c.f. Eq.~\eqref{CD_tilde_ueuo} and~\eqref{CD_tilde_delta_ueuo}). 
Due to the asymptotic behavior of Eq.~\eqref{assDeltaC}, $\Re\mathcal{K}_n < 0$ while $-\omega_I >0$, and it is the competition between these two terms that dictates whether the instability occurs in the low-lying modes.
For $2\omega_R +\Im\mathcal{K}_n>0$, the spiral occurs in the counter-clockwise direction.
For the above example regarding Eq.~\eqref{fundamentalInsPT}, we have $\mathcal{K}_n=-2\kappa$.

We close this section by elaborating on the case where the QNMs in question are attributed to the $D$'s zeros.
Once more, the modification to the Wronskian is dominated by two familiar contributions.
Specifically, we have
\bqn
\Delta W_C \doteq W\left(g,\Delta Ce^{i\omega x}\right) = \left.\frac{d\left(\Delta C\right)}{dx}\right|_{x=x_\mathrm{cut}} e^{2i\omega x_\mathrm{cut}} .
\eqn
It is noted that both $C$ and $\Delta C$ are now finite, but only $\Delta C$ will give a small contribution due to its radial dependence.
The second contribution comes from the coefficient $D$ 
\bqn
\Delta W_D \doteq W\left(g,De^{-i\omega x}\right) = -2i\omega D .
\eqn
The above contribution to the Wronskian vanishes when $D$ attains a zero, but it gives a small contribution when the frequency slightly deviates from $\omega_n$.
Again, one requires that the above two contributions precisely cancel out, giving rise to an equation determining the frequency's deviation.

Instead of expanding around the residual, one performs Taylor expansion near the zeros of $D$, namely, 
\bqn
D \equiv D(\omega) = D(\omega_n) + (\omega-\omega_n)D'(\omega_n) = (\omega-\omega_n)D'(\omega_n) ,\label{DexpZero}
\eqn
where $D'(\omega_n)$ is the derivative on the complex frequency plane, evaluated at the quasinormal frequency.

We finally arrive at a familiar form
\bqn
\delta\omega_n^{\mathrm{zero}} =\mathcal{J}_n e^{\left(-2\omega_I+\Re\mathcal{K}_n\right) x_\mathrm{cut}} e^{i\left(2 \omega_R + \Im\mathcal{K}_n\right) x_\mathrm{cut}} ,\label{fundamentalInsGenZero}
\eqn
where
\bqn
\mathcal{J}_n &=& \frac{-i}{2\omega_n D'(\omega_n)} ,\nb\\
\mathcal{K}_n &=& \frac{1}{x_\mathrm{cut}}\ln\left.\left[\frac{d}{dx}\Delta C(\omega_n)\right]\right|_{x=x_\mathrm{cut}} .
\eqn
The discussions about the spiral can be similarly carried out.
The next section will give an explicit example involving zero of $D$.

\section{Instability in the disjointed effective potentials}\label{sec3}

In the last section, the perturbation to the effective potential has been simplified to a simple ``cut'' at $x_\mathrm{cut}$.
In this section, we explore this aspect further by generalizing the form of perturbation to an arbitrary potential barrier.
Similar to Sec.~\ref{sec2}, we start with a simplified scenario that was explicitly addressed in Ref.~\cite{agr-qnm-instability-15} and also by other authors~\cite{agr-qnm-echoes-22, agr-qnm-lq-matrix-06, agr-qnm-instability-18}.
The simplified model features two disjointed square potential barriers.
For this scenario, we explicitly show that the instability readily occurs in the fundamental mode.
Subsequently, we show that the result can be straightforwardly generalized to arbitrary disjointed potential barriers.

\subsection{The spiral of low-lying QNMs in two disjointed square barriers}\label{sec3.1}

We consider two disjointed square potential barriers.
The main barrier mimics the black hole's effective potential, and a minor barrier is placed away from it, representing a minor perturbation to the former.
Let us denote the height and width of the main potential barrier by $V_1$ and $d_1$ and place it at the origin $x=0$.
The minor potential will be placed at $x=x_\mathrm{step} \gg 1$ with height and width $V_2\ll V_1$ and $d_2 \ll x_\mathrm{step}$.
Specifically, the effective potential reads~\cite{agr-qnm-instability-15, agr-qnm-lq-matrix-06, agr-qnm-instability-18}
\bqn
\widetilde{V}_\mathrm{Step}=
\left\{\begin{array}{cccc}
0     &\ \ \ \ \ & x < 0 & \ \ \mathrm{region}\ 1  \cr\\
V_1     &\ \ \ \ \ &  0 \le x \le d_1  & \ \ \mathrm{region}\ 2 \cr\\
0     &\ \ \ \ \ &  d_1 < x < x_\mathrm{step} & \ \ \mathrm{region}\ 3 \cr\\
V_2     &\ \ \ \ \ &  x_\mathrm{step} \le x \le x_\mathrm{step}+d_2 & \ \ \mathrm{region}\ 4  \cr\\
0     &\ \ \ \ \ &  x > x_\mathrm{step}+d_2  & \ \ \mathrm{region}\ 5
\end{array}\right. .
\label{V_spb}
\eqn

Due to the simplicity of the above model, one can evaluate the Wronskian using explicit forms of the in-going and out-going waves.
The in-going wave $f$ in region 3 possesses the same form as Eq.~\eqref{f_mPT_CD}
\bqn
f(\omega,x) = C e^{i\omega x} + D e^{-i\omega x} ,
\eqn
where the coefficients $C$ and $D$ can be evalulated straightforwardly and given by Eqs.~\eqref{CDstep} in Appx.~\ref{appB}.
Using these explicit forms, it is not difficult to verify that this particular example belongs to the second class of models discussed in Sec.~\ref{sec2.2} for which the QNMs are governed by the zero of $D$~\cite{agr-qnm-lq-matrix-06} whereas these frequencies do not coincide with any of $C$'s poles~\footnote{For the particular scenario discussed in this subsection, apart from the common factor $\omega$ on the denominators of $C$ and $D$ which is physically irrelevant, we note that $\omega=-i\sqrt{V}$ is not a pole of $C$.}.
The out-going wave $g$ in region 3 can be obtained in a similar fashion
\bqn
g(\omega,x) = E e^{i\omega x} + F e^{-i\omega x} ,
\eqn
where the coefficients $E$ and $F$ are given by Eqs.~\eqref{EF_step}.

It is important to note that the information about the location of the perturbation $x_\mathrm{step}$ is carried by the coefficient $F$ through the exponential factor $e^{2i \omega x_\mathrm{step}}$.
Also, it is readily verified that at the limit of small perturbation $V_2\to 0$, we have 
\bqn
\lim\limits_{V_2\to 0} E = 1, \label{Elim}
\eqn
and 
\bqn
\lim\limits_{V_2\to 0} \frac{F}{V_2} = e^{2i\omega x_\mathrm{step}}e^{i\omega d_2}\frac{i\sin\left(\omega d_2\right)}{2\omega^2}. \label{Flim}
\eqn
The limits given by Eqs.~\eqref{Elim} and~\eqref{Flim} are physical.
As $V_2\to 0$, the reflection amplitude vanishes along with the size of the potential, while the transmission approaches 100\%. 

The Wronskian gives
\bqn
W(g, f)= 2i\omega (CF - DE) ,
\eqn
which can be further elaborated by considering the deviation from a given low-lying QNM $\omega_n$. 
Similar to Sec.~\ref{sec2.2}, one can approximate $D$ by its first-order derivative as given by Eq.~\eqref{DexpZero}.
It is observed that the smallness of the deviation $\delta\omega_n=\omega-\omega_n$ is to be matched to the size of $F$ given by the limit Eq.~\eqref{Flim}.
Specifically, we have 
\bqn
\delta\omega_n^{\mathrm{zero}} =\mathcal{J}_n e^{-2\omega_I x_\mathrm{step}} e^{2i \omega_R x_\mathrm{step}} ,\label{DoubleSqureBarrier}
\eqn
where
\bqn
\mathcal{J}_n = \frac{V_2 C(\omega_n)}{D'(\omega_n)}e^{i\omega d_2}\frac{i\sin\left(\omega_n d_2\right)}{2\omega_n^2} \label{JnDisjntPotBar}.
\eqn
The discussions about the spiral can be similarly carried out.
Notably, in this case, the fundamental mode is always unstable due to the absence of the term $\mathcal{K}_n$ (c.f. Eq.~\eqref{fundamentalInsGenZero}), and the spiral always rotates in the counter-clockwise direction\footnote{For the two examples given in Ref.~\cite{agr-qnm-instability-15}, the spiral occurs in the clockwise direction since the vertical axis is shown in $-\omega_I$, instead of $\omega_I$.}.

Also, we note that at the limit $d_2\to 0$ while keeping $V_2$ a finite value, the deviation Eq.~\eqref{DoubleSqureBarrier} vanishes.
This is readily verified by observing Eq.~\eqref{JnDisjntPotBar} as the amplitude of the spiral becomes zero.
The result for this limit is intuitive, as a potential barrier with a vanishing width will not interfere with wave propagation.

\subsection{The generalization from disjointed to continuous perturbation of an arbitrary form}\label{sec3.2}

The generalization of the above result to two arbitrary disjointed potential barriers is somewhat straightforward, which will be carried out in this subsection.
Moreover, by comparing the physical scenario elaborated in Sec.~\ref{sec2}, we will further generalize earlier results to consider a scenario where a perturbative potential barrier of arbitrary form is placed on top of a continuous black hole effective potential.

First, we note that for two disjointed potentials of arbitrary form, it is relatively intuitive that the specific forms of the coefficients $C, D, E$, and $F$ are not crucial in deriving Eqs.~\eqref{DoubleSqureBarrier}.
Specifically, one only needs the expansion Eq.~\eqref{DexpZero} to account for the contribution near the zero and the limiting behavior of the amplitudes Eq.~\eqref{Elim},
\bqn
F = e^{2i\omega x_\mathrm{step}}\widetilde{F}, \label{FFtilde}
\eqn
and
\bqn
\lim\limits_{V_2\to 0}\widetilde{F} = 0, \label{FlimMod}
\eqn
to take into account the magnitude of $\delta\omega_n$, where $\lim\limits_{V_2\to 0}\frac{\widetilde{F}}{V_2}$ is a finite well-defined quantity and independent of either $x_\mathrm{step}$ and overall strength of the perturbation.

One find, subsequently,
\bqn
\delta\omega_n^{\mathrm{zero}} =\mathcal{J}_n e^{-2\omega_I x_\mathrm{step}} e^{2i \omega_R x_\mathrm{step}} ,\label{DoubleBarrier}
\eqn
where
\bqn
\mathcal{J}_n = \frac{C(\omega_n)\widetilde{F}(\omega_n)}{D'(\omega_n)} .
\eqn
Therefore, one concludes that the fundamental mode is intrinsically unstable given that the black hole's effective potential and the perturbation are defined on separated compact domains\footnote{After we submitted the manuscript, we became aware of the studies~\cite{agr-qnm-instability-56, agr-qnm-instability-58}, where similar arguments have been employed to address the instability in the fundamental mode for disjoint potential barriers.}.

One may pursue this line of reasoning further and aim to explore a more general scenario: a perturbative potential barrier is placed near spatial infinity but on top of a realistic black hole effective potential.
To this end, we inspect the physical origin that leads to the difference between Eqs.~\eqref{DoubleBarrier} and~\eqref{fundamentalInsGenZero}.
The answer resides in the asymptotic behavior of the effective potential at spatial infinity.
In deriving Eq.~\eqref{fundamentalInsGenZero}, the contribution of $\mathcal{K}_n$ comes from the correction $\Delta C$, which vanishes identically in the case of disjointed potentials given by Eq.~\eqref{DoubleBarrier}.
For the latter, one observes that the black hole effective potential is suppressed essentially by an exponential form, such as the P\"oschl-Teller potential Eq.~\eqref{V_PT}.
Moreover, it is well known for a more realistic case that the Regge-Wheeler potential is suppressed at spatial infinity by merely a power-law form Eq.~\eqref{V_RW}.
In other words, the contribution from the correction $\Delta C$ is larger and, therefore, the simplification introduced in deriving Eq.~\eqref{DoubleBarrier} must be rectified cautiously.
On the other hand, by considering a potential barrier of arbitrary form, one picks up an additional minor in-going component for the waveform $g(\omega, x)$, carried by $F$, which was entirely ignored in deriving Eq.~\eqref{fundamentalInsGenZero}.

In the above derivations, either the truncation $x_\mathrm{cut}$ or the potential barrier $x_\mathrm{step}$ denotes a characteristic length scale of the system, embedded in either $\Delta C$ or $F$, that gives rise to an exponential form.
Mathematically, it can be viewed as a spatial translation operation.
This factor is responsible for the spiral feature in the small deviation from the original quasinormal frequency.
Besides, the smallness of the deviation resides in Eq.~\eqref{assDeltaC} or~\eqref{FlimMod}.
By recuperating the contribution from $\Delta C$ from Eq.~\eqref{fundamentalInsGenZero} and plugging it into Eq.~\eqref{DoubleBarrier}, we now take into account both factors and find
\bqn
W(g, f)\doteq 2i\omega (CF - D)+ \left.\frac{d\Delta C}{dx}\right|_{x=x_\mathrm{step}}e^{2i\omega x_\mathrm{step}} ,
\eqn
which subsequently gives
\bqn
\delta\omega_n =\mathcal{J}_n e^{\left(-2\omega_I+\Re\mathcal{K}_n\right) x_\mathrm{step}} e^{i\left( 2\omega_R+\Im\mathcal{K}_n\right) x_\mathrm{step}} ,\label{SpiralZeroPoleGen}
\eqn
where
\bqn
\mathcal{J}_n &=& \frac{C(\omega_n)\widetilde{F}(\omega_n)}{D'(\omega_n)} ,\nb\\
\mathcal{K}_n &=& \frac{1}{x_\mathrm{cut}}\ln\left.\left[1-\frac{i}{2\omega_n C(\omega_n)\widetilde{F}(\omega_n)}\frac{d}{dx}\Delta C(\omega_n)\right]\right|_{x=x_\mathrm{step}} .\label{JKzerosGen}
\eqn
Alternatively, when the QNM is attributed to a pole in the transmission amplitude $C$, 
we have
\bqn
\frac1C W(g, f)\doteq 2i\omega \left(F - \frac{D}{C}\right)+ \left.\frac{d}{dx}\left(\frac{\Delta C}{C}\right)\right|_{x=x_\mathrm{step}}e^{2i\omega x_\mathrm{step}} .
\eqn
Subsequently, the relevant quantities in Eq.~\eqref{SpiralZeroPoleGen} defined by
\bqn
\mathcal{J}_n &=& \frac{\mathrm{Res}\left(C,\omega_n\right)\widetilde{F}(\omega_n)}{D(\omega_n)} ,\nb\\
\mathcal{K}_n &=& \frac{1}{x_\mathrm{cut}}\ln\left.\left[1-\frac{i}{2\omega_n \widetilde{F}(\omega_n)}\frac{d}{dx}\lim\limits_{\omega\to\omega_n}\frac{\Delta C}{C}\right]\right|_{x=x_\mathrm{step}} .\label{JKpolesGen}
\eqn
Eqs.~\eqref{SpiralZeroPoleGen},~\eqref{JKzerosGen}, and~\eqref{JKpolesGen} summarize the results obtained by assuming the most general context.

\section{Numerical calculations}\label{sec4}

Now, we are in a position to present the numerical results. 
We first present the results for the modified P\"oschl-Teller effective potential elaborated in Sec.~\ref{sec2.1}.
To facilitate the numerical calculations, we consider the following stepwise effective potential
\bqn
\widetilde{V}_\mathrm{PT}=
\left\{\begin{array}{cc}
{V}_\mathrm{PT}(V_{m-},\kappa_-, x)     &  x\le x_\mathrm{step}  \cr\\
{V}_\mathrm{PT}(V_{m+},\kappa_+, x)    &  x> x_\mathrm{step}  
\end{array}\right. ,
\label{V_mpt_mod}
\eqn
where the difference in the parameters introduces a minor discontinuity at $x=x_\mathrm{step}$.
The different potential form for the region $x> x_\mathrm{step}$ will not significantly affect the derivation given in Sec.~\ref{sec2.1}.
Indeed, it will modify the out-going waveform Eq.~\eqref{g_mPT} by including a further correction, which readily vanishes as $x\to +\infty$.
As a result, it will modify the specific form of $D(\omega_n)$ on the denominator of Eq.~\eqref{JformPole1} but will not impact any essential conclusion about the spiral features.

\begin{figure}[htp]
\centering
\includegraphics[scale=0.54]{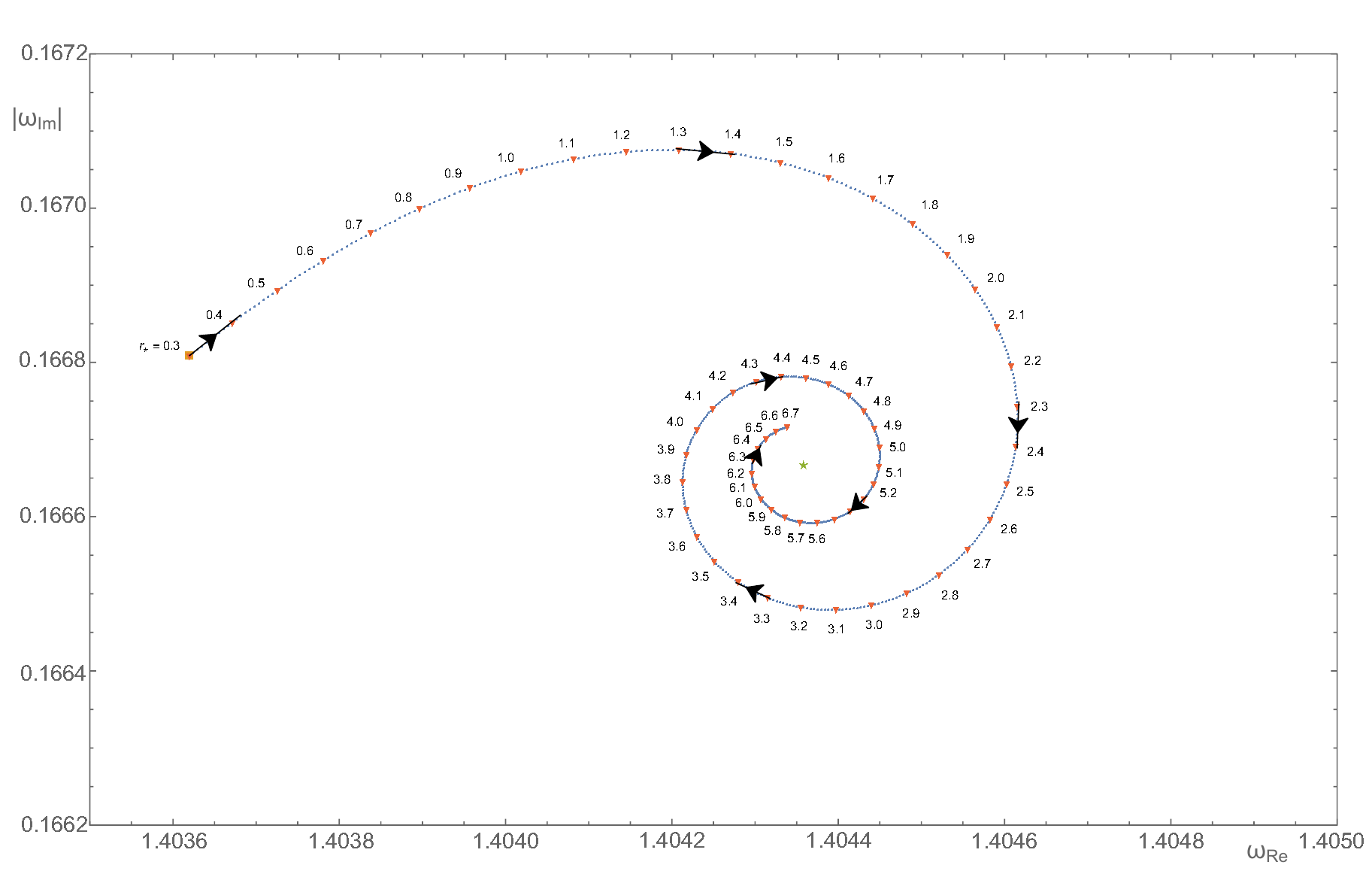}
\caption{Evolution of the fundamental mode as $x_\mathrm{step}$ moves outward. 
The calculations are carried out by using the effective potential Eq.~\eqref{V_mpt_mod} with the parameters $\kappa_{-}=\kappa_{+}=1/3$ and $V_{m-}=2,V_{m+}=1.995$.
The green filled star corresponds to the mode of the original P\"oschl-Teller effective potential.
The fundamental mode is manifestly stable as it spirals inward as the perturbation moves away from the potential.}
\label{fig1}
\end{figure}

\begin{figure}[htp]
\centering
\includegraphics[scale=0.54]{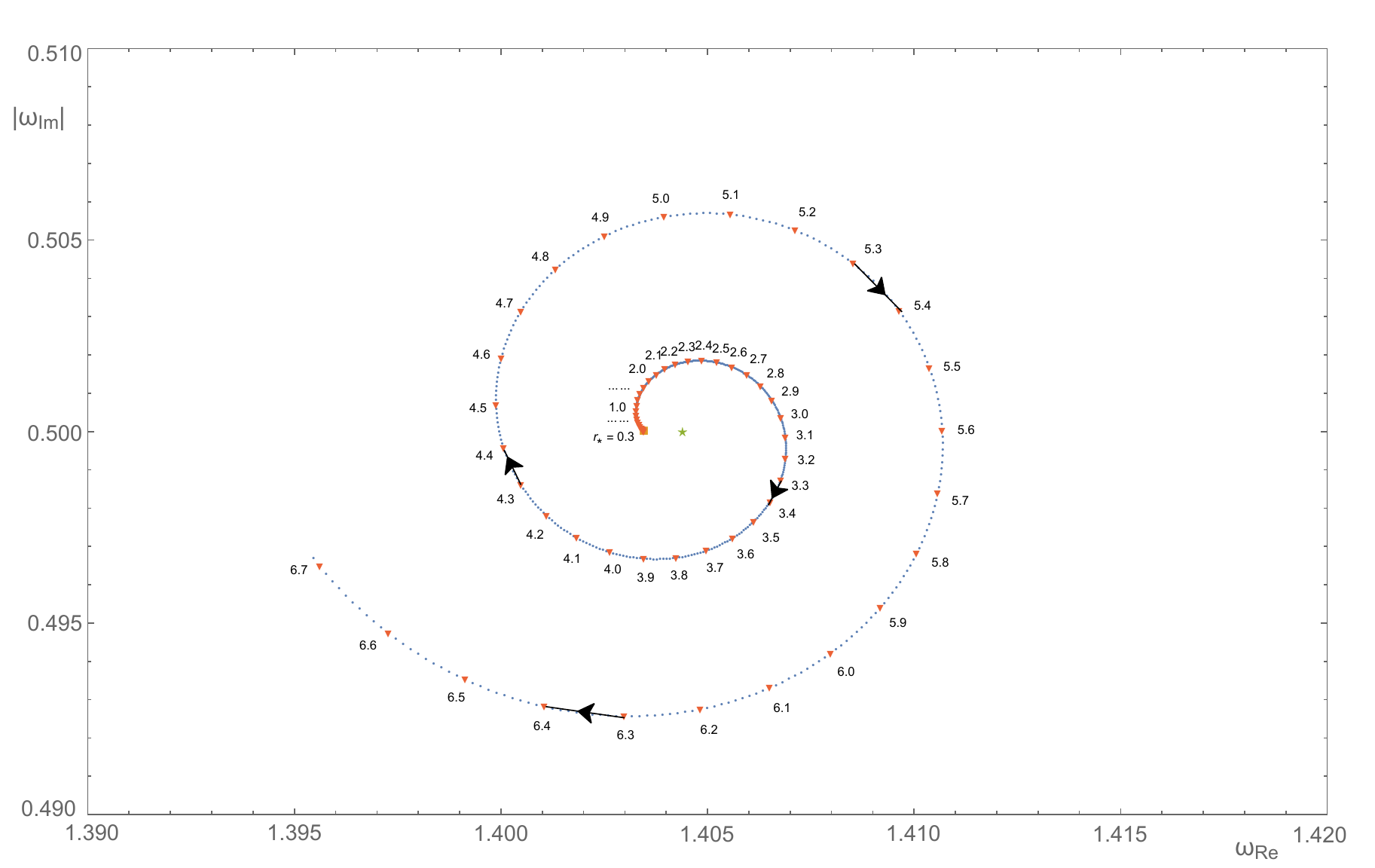}
\caption{The same as Fig.~\ref{fig1} but for the first overtone.
Unlike the fundamental mode, the first overtone is unstable as it spirals outward as the perturbation moves away from the potential.}
\label{fig2}
\end{figure}

\begin{figure}[htp]
\centering
\includegraphics[scale=0.54]{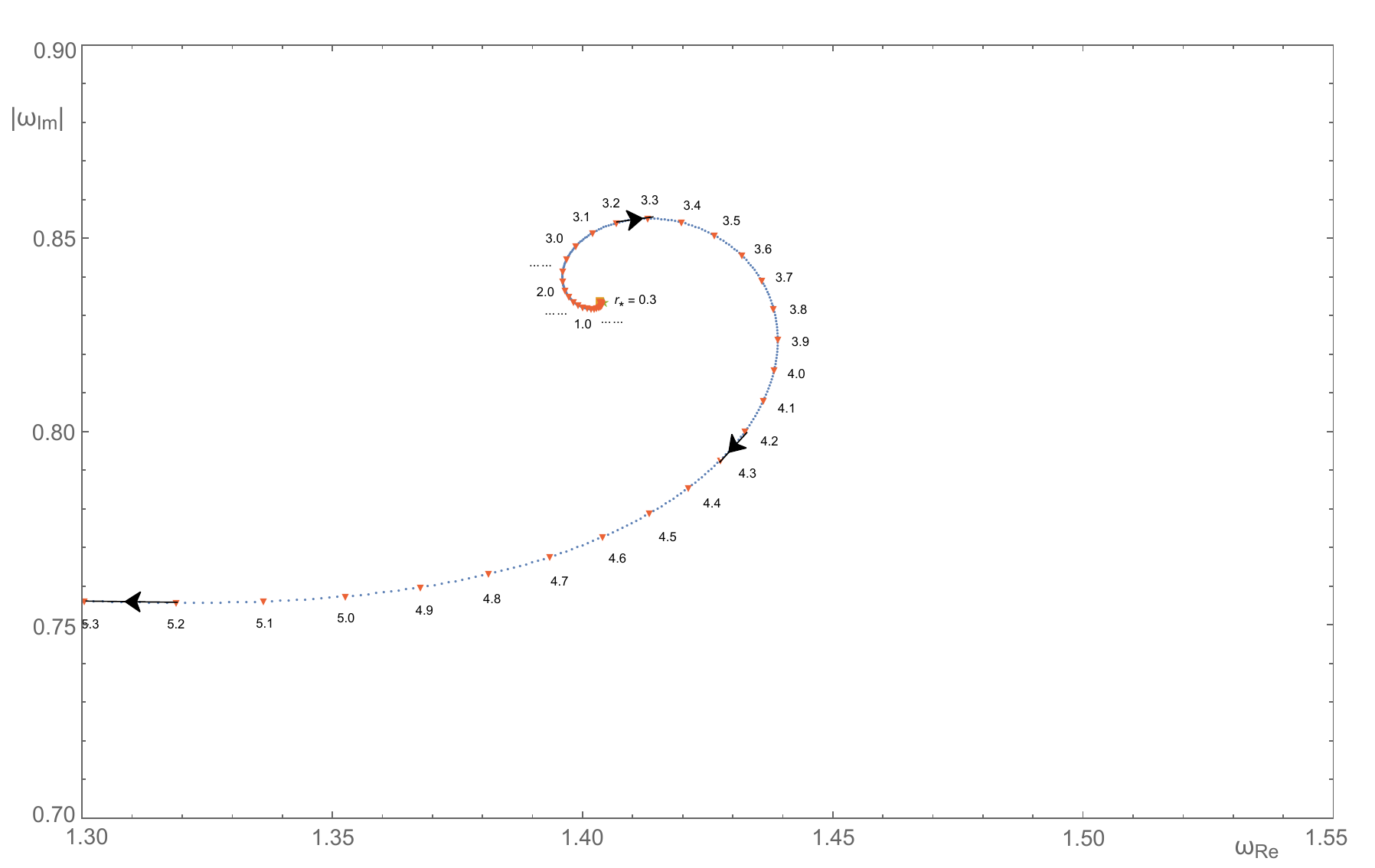}
\caption{The same as Fig.~\ref{fig1} but for the second overtone.
It is also found to be unstable as it spirals outward as the perturbation moves away from the potential.}
\label{fig3}
\end{figure}

The evolution of the fundamental, first, and second overtones in the complex frequency plane as functions of $x_\mathrm{step}$ are presented in Figs.~\ref{fig1},~\ref{fig2} and~\ref{fig3}.
The numerical calculations are carried out using the matrix method~\cite{agr-qnm-lq-matrix-02, agr-qnm-lq-matrix-04, agr-qnm-lq-matrix-11, agr-qnm-lq-matrix-12}.
The obtained results have been verified by explicitly evaluating the Wronskian Eq.~\eqref{pt_Wronskian}, as in this particular case, the QNMs are governed by the zeros.

It is apparent that the fundamental mode spirals inward, manifesting its stability, while the remaining overtones spiral outward, indicating their instabilities.
We also note that all the modes rotate in the counter-clockwise direction, though in the plots, the imaginary axis is flipped, and therefore, visually, they rotate in the clockwise direction.
Now, one proceeds to compare the obtained spiral periods and the deviations from the original QNMs extracted from the numerical calculations against the analytic estimation given by Eq.~\eqref{fundamentalInsPT}.
According to Eq.~\eqref{fundamentalInsPT}, the spiral period is universally given by $\Delta x_\mathrm{step} =\pi/\sqrt{V_{m}-\kappa^2/4} \simeq 2.24$.
This can be compared with the numerical results.
In Fig.~\ref{fig1}, the displacement of $x_\mathrm{cut}$ for the innermost spiral of the fundamental mode is $\Delta x_0 \simeq 6.7-4.4 = 2.3$, and the second spiral gives $\Delta x_0 \simeq 4.4-1.5=2.9$.
In Fig.~\ref{fig2}, the innermost spiral of the first overtone is $\Delta x_1 \simeq 4.4-2.0=2.4$, and the second spiral gives $\Delta x_1 \simeq 6.7-4.4=2.3$.
In Fig.~\ref{fig3}, as the trajectory has not completed an entire cycle, one roughly estimates the spiral period by a half-circle as $\Delta x_2 \simeq 2(4.2-3.0)=2.4$.
Moreover, we can compare the deviation from the original location of the QNM as a function of $x_\mathrm{step}$.
Reasonable agreement is observed in Tab.~\ref{Tab.1}, where the analytic estimation is given by $e^{(2n-1)\kappa \Delta x_\mathrm{step}}$ due to Eq.~\eqref{fundamentalInsPT}.
One, therefore, concludes that the analytic estimation given by Eq.~\eqref{fundamentalInsPT} is sound.

\begin{table}[htbp] 
\caption{\label{Tab.1} 
The relative deviations of the fundamental frequency $\left|\delta\omega_0(x_\mathrm{step})/\delta\omega_0(x_\mathrm{step}=1.8)\right|$ as a function of $x_\mathrm{step}$ for the modified P\"oschl-Teller effective potential.} 
\adjustbox{max width=\textwidth}{%
\begin{tabular}{|c|cccccccccc|}
\hline
          $x_\mathrm{step}$ &$1.8$&$2.0$&$2.2$&$2.4$&$2.6$&$2.8$&$3.0$&$3.2$&$3.4$&$3.6$  \\
\hline
    analytic estimation&$1$&$~0.94~$&$~0.88~$&$~0.82~$&$~0.77~$&$~0.72~$&$~0.67~$&$~0.63~$&$~0.59~$&$~0.55~$  \\
    numerical result&$1$&$~0.91~$&$~0.83~$&$~0.76~$&$~0.70~$&$~0.64~$&$~0.59~$&$~0.54~$&$~0.50~$&$~0.49~$  \\
\hline
\end{tabular}}
\end{table}

As a second example, we consider the scenario where the black hole metric and the perturbations are disjointed, as elaborated in Sec.~\ref{sec3.1}.

\begin{figure}[htp]
\centering
\includegraphics[scale=0.418]{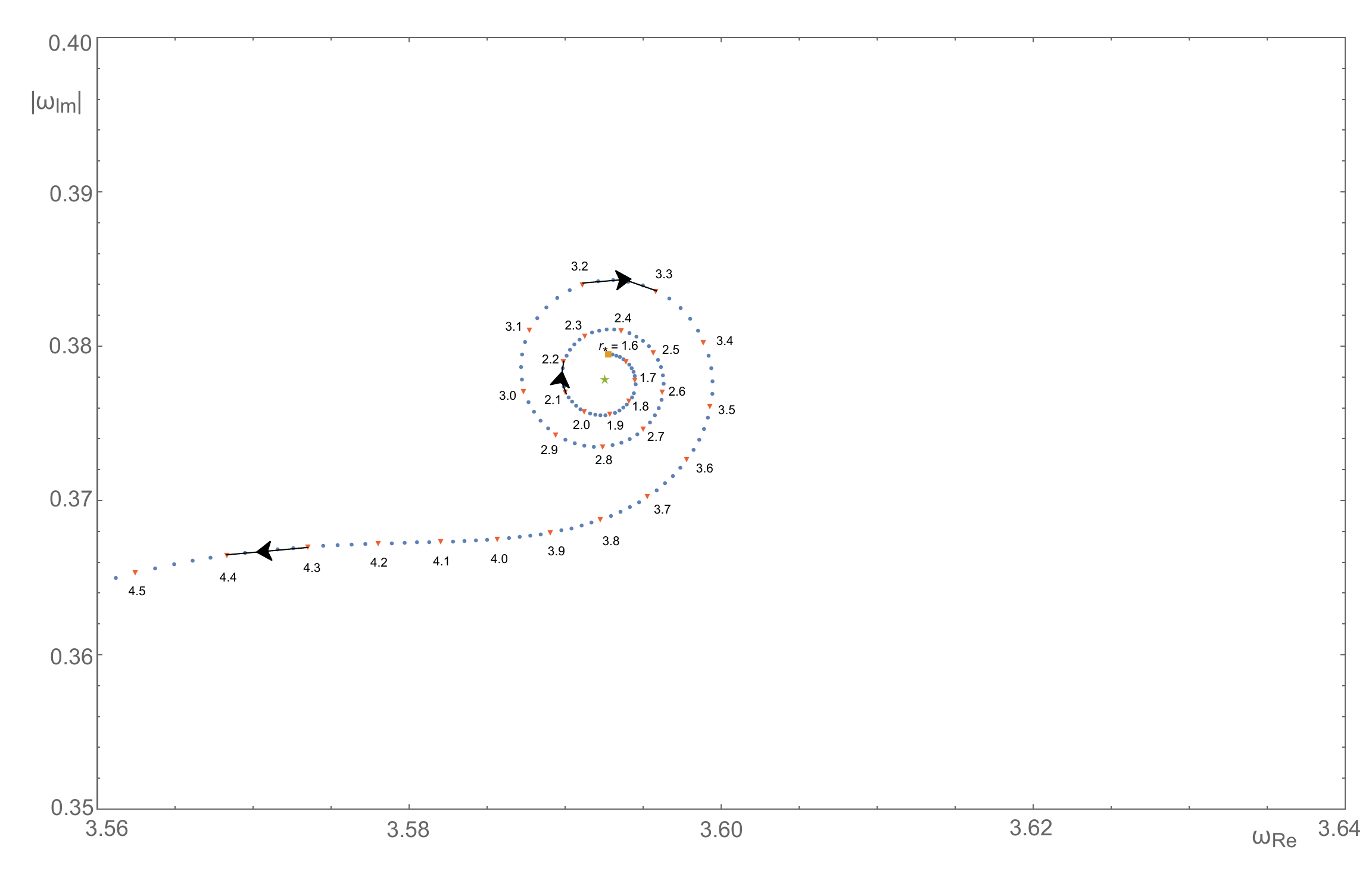}
\caption{Evolution of the fundamental mode as $x_\mathrm{step}$ moves outward. 
The calculations are carried out using the effective potential Eq.~\eqref{V_spb} with the parameters $d_1=d_2=1.5$, $V_1=10$, and $V_2=0.05$.
The green filled star corresponds to the mode of the unperturbed rectangle effective potential.
The fundamental mode is unstable as it spirals outward as the perturbation $V_2$ moves away from the primary potential $V_1$.}
\label{fig4}
\end{figure}

\begin{figure}[htp]
\centering
\includegraphics[scale=0.42]{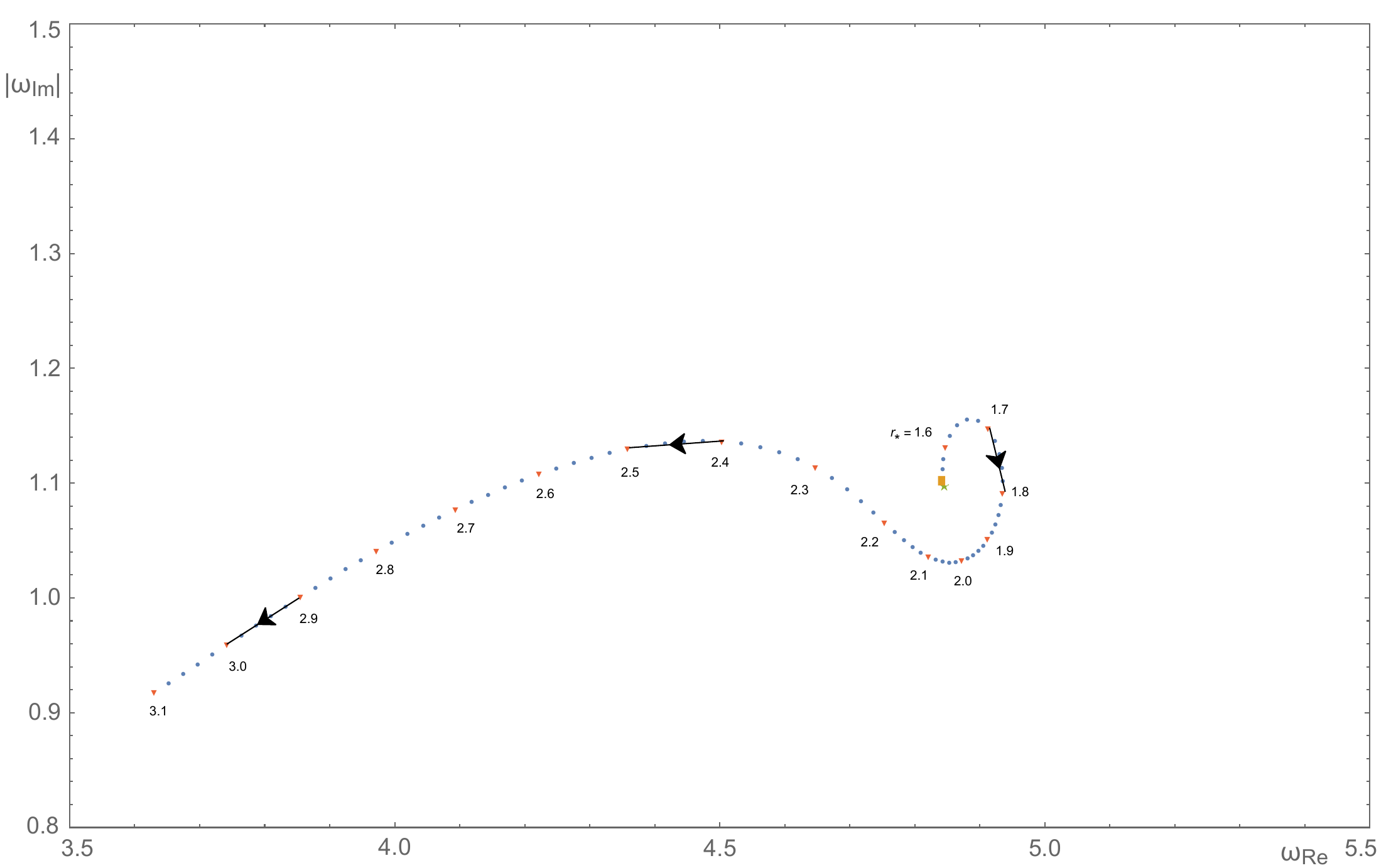}
\caption{The same as Fig.~\ref{fig4} but for the first overtone.
The first overtone is also unstable as it spirals outward as the perturbation $V_2$ moves away from the primary potential $V_1$.}
\label{fig5}
\end{figure}

For the effective potential Eq.~\eqref{V_spb}, we evaluate the evolution of the fundamental mode and first overtone, as shown in Fig.~\ref{fig4} and~\ref{fig5}.
Numerical calculations for such rectangular potential barriers have already been presented in~\cite{agr-qnm-instability-15, agr-qnm-instability-32}.
Here, again, let us quantitatively compare the obtained spiral periods and the deviations from the original QNMs extracted from the numerical calculations with the analytic estimation given by Eq.~\eqref{DoubleSqureBarrier}.
The latter indicates that the spiral period is governed by the real parts of the QNMs that pertain to the unperturbed effective potential. 
The first two QNMs of the primary rectangular potential for the adopted parameters read $\omega_0 = 3.5925368431304348096 - 0.37781367044298397773 i$ and $\omega_1 = 4.8783589468037452649 - 1.0961146541752961631 i$. 
Therefore, it is expected from Eq.~\eqref{DoubleSqureBarrier} that the spiral period near the fundamental mode is approximately $\Delta x_\mathrm{step} =\pi/\Re\omega_0 \simeq 0.87$, and the period for the first overtone is $\Delta x_\mathrm{setp} = \pi/\Re\omega_1 \simeq 0.64$.
From Fig.~\ref{fig4}, one can numerically extract the period near the fundamental mode from the innermost spiral, which gives ${\Delta} x_{0}\simeq 2.46-1.6 = 0.86$.
Also, for the second spiral gives ${\Delta} x_{0}\simeq 3.34-2.46=0.88$.
Similarly, from Fig.~\ref{fig5}, one finds that the spiral period of the first overtone is approximately ${\Delta} x_{1}\simeq 2.28-1.6=0.68$. 
Besides, we can also compare the deviation from the original location of the QNM as a function of $x_\mathrm{step}$.
This is shown in Tab.~\ref{Tab.2}, where the relative deviation is given analytically by $e^{-2\omega_I \Delta x_\mathrm{step}}$ due to Eq.~\eqref{DoubleSqureBarrier}.
These comparisons demonstrate that the analysis given in the previous sections is qualitatively valid.

\begin{table}[htbp] 
\caption{\label{Tab.2} The relative deviations of the fundamental frequency $\left|\delta\omega_0(x_\mathrm{step})/\delta\omega_0(x_\mathrm{step}=1.8)\right|$ as a function of $x_\mathrm{step}$ for the effective potential consisting of two disjointed square barriers.} 
\adjustbox{max width=\textwidth}{%
\begin{tabular}{|c|cccccccccc|}
\hline
          $x_\mathrm{step}$ &$1.8$&$2.0$&$2.2$&$2.4$&$2.6$&$2.8$&$3.0$&$3.2$&$3.4$&$3.6$  \\
\hline
    analytic estimation&$1$&$~1.16~$&$~1.35~$&$~1.57~$&$~1.83~$&$~2.13~$&$~2.48~$&$~2.88~$&$~3.35~$&$~3.90~$  \\
    numerical result&$1$&$~1.17~$&$~1.39~$&$~1.63~$&$~1.82~$&$~2.08~$&$~2.52~$&$~3.05~$&$~3.27~$&$~3.54~$  \\
\hline
\end{tabular}}
\end{table}

Before closing this section, it is worth mentioning some subtleties between the above examples and more realistic scenarios.
First, for asymptotically flat spacetime, it is expected that the size of the discontinuity or height of the bump asymptotically decreases as $1/r^2$, which ensures that the mass associated with the metric perturbation is roughly a constant.
One can show that such a nuance will not change the main conclusion from both analytic and numerical sides.
Taking the stepwise P\"oschl-Teller potential Eq.~\eqref{V_mpt_mod} as an example, the derivation will not differ significantly from those presented above.
When comparing to the truncated effective potential, the out-going waveform Eq.~\eqref{g_mPT} will contain a small fraction of ingoing wave, which is primarily suppressed exponentially as $x_\mathrm{step}\to\infty$.
As discussed above, although it will modify the specific form of $D(\omega_n)$ on the denominator of Eq.~\eqref{JformPole1}, it will not impact any essential conclusion about the spiral features in terms of Eq.~\eqref{fundamentalInsPT}.
In addition, we have verified that numerical results presented in Fig.~\ref{fig1} essentially stay unchanged when one modifies the value of $V_{m+}$ so that $\Delta V\equiv V_{m-}-V_{m+}\propto 1/x^2$.

Moreover, regarding the asymptotical behavior of the effective potential, the Regge-Wheeler potential decays slowly as $1/r^2$.
Meanwhile, both of the explicitly calculated examples decay faster. Specifically, the P\"oschl-Teller effective potential decays exponentially while the disjoint potential exaggerates the scenario further and ignores the potential entirely at large spatial coordinates.
These different asymptotical behaviors warrant further analysis if one wishes to extend the obtained analytical results to the Regge-Wheeler potential.
However, it is rather challenging to generalize the analytic calculations performed in this study to the case of Regge-Wheeler potential.
This is because, unfortunately, we do not possess sufficient information on either the analytically close forms or asymptotical approximations of the wave functions.
Moreover, to one's surprise, one finds a seeming contradiction by looking closer at the obtained stability of the fundamental mode.
Specifically, in the case of disjoint potential barriers, which decrease the fastest, the fundamental mode is manifestly unstable.
For the case of modified P\"ochl-Teller potential, which decreases slower but still faster than the Regge-Wheeler one, it is shown that the fundamental mode is stable.
Therefore, it is not straightforward to understand why the Regge-Wheeler potential, as shown by existing studies, is featured by an unstable fundamental mode.
This indicates that the specific form of the effective potential, and therefore the resulting waveform, is crucial, besides the fact that the asymptotic behavior of the effective potential must also play a pivotal role.
This gives rise to an intriguing problem currently lying beyond our reach.
Nonetheless, calculations can be achieved numerically for Regge-Wheeler potential with a metric perturbation featuring a discontinuity via a step or truncation.
Numerical results indicate instability in the fundamental modes, consistent with findings in~\cite{agr-qnm-instability-15}.

\section{Concluding remarks}\label{sec5}

To summarize, we explore the instability in the low-lying QNMs from an analytic perspective.
We start with a specific case of P\"oschl-Teller potential, where the deviation from an original QNM is assessed by evaluating the residual around a pole of the Gamma function.
Proceeding further, the universality of the phenomenon is elaborated regarding two different mathematical origins of the QNMs: the poles in the reflection amplitude and zeros in the transmission amplitude.
The perturbation is planted further away from the black hole's horizon and can be simplified to a discontinuity in the effective potential.
The discussions then extended to consider perturbations that have an arbitrary finite shape.
The derivations are accomplished by first considering a simplified case where perturbations are defined on a separated compact domain and then generalizing it to a continuous scenario. 
The numerical results for the spiral period and the relative deviation of the quasinormal frequency align well with our predictions, further validating the analytical derivations.
The resulting instability has the following main features
\begin{itemize}
    \item The phenomenon is generally relevant for the first few low-lying modes but might be suppressed for the lowest-lying ones. 
    A criterion for its emergence is derived and summarized in Eqs.~\eqref{SpiralZeroPoleGen},~\eqref{JKzerosGen}, and~\eqref{JKpolesGen} for a given black hole metric and specific perturbations.
    \item The spiral is more pronounced when the magnitude of the original quasinormal frequency's imaginary part is smaller than the real part.
    \item Mathematically, the phenomenon can be attributed to an interplay between the asymptotical behavior near the singularity associated with QNM and the spatial translation applied to the effective potential's perturbation.
\end{itemize}
The last point deserves some clarification.
The physical system is characterized by a length scale $x_\mathrm{cut}$ (or $x_\mathrm{step}$) owing to the spatial translation of the perturbation.
It affects the Wronskian on the denominator of the Green's function through an exponential factor.
The exponential form mainly dictates that the length scale is the dominant variable.
To assess the QNM, the above effect must, by definition, be canceled out by the frequency's deviation from its original value, whose contribution to the Wronskian is implemented through the pole's residual (or the first-order derivative at the zero).
The interplay between the above two factors gives rise to the observed spiral.

The analytic derivations performed in the present study predominantly reside in the domain of linear stability analysis.
For all the assumptions proposed in Sec.~\ref{sec1} to be valid, one demands that the metric perturbation be small enough so that the QNM's deviation can still be considered perturbative.
Sometimes, the deviation in quasinormal frequency might increase exponentially and substantially, leading to linear instability.
The latter, however, will rapidly invalidate the assumption of the smallness of the frequency deviation. 
As a result, in practice, such linear instability might not always imply global instability; for instance, the perturbed system might evolve into a stable soliton solution instead of blowing up exponentially.
In our case, however, numerical calculations indicate that our findings are rather robust.
Specifically, the deviation of the QNM can be rather sizable before the analytical estimation ceases to be valid, where the fundamental mode might be taken over by another mode.

Also, we note that the mathematical analysis carried out in this study is consistent with those performed in~\cite{agr-qnm-lq-03, agr-qnm-echoes-20}.
For the particular case of truncated P\"oschl-Teller potential, all these analyses share precisely the same junction condition, which can be implemented by a vanishing Wronskian.
Also, the specific asymptotical forms of the wavefunctions are identical.
On the one hand, to explore the deviation of the fundamental mode, we are focused on the vicinity of the original black hole's quasinormal frequencies.
The latter is utilized to determine how such a deviation evolves as a function of the location of the perturbation.
On the other hand, to assess the high overtones, one uses the fact that these modes have large real parts.
Subsequently, the asymptotical behavior of the QNMs constituting a logarithmic branch can be derived, which is closely related to the spectral instability.

On the physical side, despite recent progress, we understand that the problem is not entirely settled.
As initially pointed out by Nollert, Price, Aguirregabiria, and Vishveshwara in their seminal papers~\cite{agr-qnm-35, agr-qnm-36, agr-qnm-27, agr-qnm-30} and recently revitalized by Jaramillo {\it et al.}~\cite{agr-qnm-instability-07, agr-qnm-instability-13}, and some of us~\cite{agr-qnm-50, agr-qnm-lq-03, agr-qnm-echoes-20}, structural instability poses a substantial challenge to the black hole spectroscopy.

\section*{Acknowledgements}

We gratefully acknowledge the financial support from Brazilian agencies 
Funda\c{c}\~ao de Amparo \`a Pesquisa do Estado de S\~ao Paulo (FAPESP), 
Funda\c{c}\~ao de Amparo \`a Pesquisa do Estado do Rio de Janeiro (FAPERJ), 
Conselho Nacional de Desenvolvimento Cient\'{\i}fico e Tecnol\'ogico (CNPq), 
and Coordena\c{c}\~ao de Aperfei\c{c}oamento de Pessoal de N\'ivel Superior (CAPES).
This work is supported by the National Natural Science Foundation of China (NSFC).
A part of this work was developed under the project Institutos Nacionais de Ci\^{e}ncias e Tecnologia - F\'isica Nuclear e Aplica\c{c}\~{o}es (INCT/FNA) Proc. No. 464898/2014-5.
This research is also supported by the Center for Scientific Computing (NCC/GridUNESP) of S\~ao Paulo State University (UNESP).

\appendix

\section{Expansion coefficients for the waveforms in the truncated P\"oschl-Teller potential}\label{appA}

In this appendix, we give the specific expressions utilized in the main text regarding the problem of the (modified) P\"oschl-Teller potential.
Further details concerning the derivation can be found in the Appendix of Ref.~\cite{agr-qnm-lq-03}

The formal solutions of the Schr\"odinger equation Eq.~\eqref{pt_homo_eq} with the potential $U=-{V}_\mathrm{PT}$ can be found in standard textbooks such as Ref.~\cite{book-quantum-mechanics-Flugge}.
One may conveniently choose two independent solutions with even and odd parities, namely, $u_e(-x)=u_e(x)$ and $u_o(-x)=-u_o(x)$, which read
\bqn
u_e(x) &=&\cosh^\lambda \kappa x {_2F_1}\left(a,b,\frac12;-\sinh^2 \kappa x\right) ,\nonumber\\
u_o(x) &=&\cosh^\lambda \kappa x \sinh \kappa x {_2F_1}\left(a+\frac12,b+\frac12,\frac32;-\sinh^2 \kappa x\right) ,
\label{ueuo_solution}
\eqn
where
\bqn
a = \frac12\left(\lambda+i\frac{\omega}{\kappa}\right) ,\nonumber\\
b = \frac12\left(\lambda-i\frac{\omega}{\kappa}\right) ,
\label{def_ab}
\eqn
and
\bqn
\lambda =  \frac12 + \frac{\sqrt{4V_m+\kappa^2}}{2\kappa} .
\label{def_lambda}
\eqn
It is noted that $\lambda > 1$ when $V_m$ and $\kappa$ are positive real numbers.

We aim at an appropriate combination defined in Eq.~\eqref{f_mPT}, which agrees with the boundary condition given in the first line of Eq.~\eqref{pt_boundary}.
By using the expansion formulae of ${_2F_1}(a,b,\frac12;z)$ and ${_2F_1}(a,b,\frac32;z)$ at $z\to \infty$ to the first order, one has
\bqn
u_e(x) &\to& \Gamma\left(\frac12\right)\left\{\frac{\Gamma\left(\frac{-i\omega}{\kappa}\right)e^{i\frac{\omega}{\kappa}\ln 2}}{\Gamma\left(\frac{\lambda}{2}-i\frac{\omega}{2\kappa}\right)\Gamma\left(\frac{1-\lambda}{2}-i\frac{\omega}{2\kappa}\right)}e^{i\omega x}+
\frac{\Gamma\left(\frac{i\omega}{\kappa}\right)e^{-i\frac{\omega}{\kappa}\ln 2}}{\Gamma\left(\frac{\lambda}{2}+i\frac{\omega}{2\kappa}\right)\Gamma\left(\frac{1-\lambda}{2}+i\frac{\omega}{2\kappa}\right)}e^{-i\omega x}\right\} ,\nonumber\\
u_o(x) &\to& -\Gamma\left(\frac32\right)\left\{\frac{\Gamma\left(\frac{-i\omega}{\kappa}\right)e^{i\frac{\omega}{\kappa}\ln 2}}{\Gamma\left(\frac{\lambda+1}{2}-i\frac{\omega}{2\kappa}\right)\Gamma\left(\frac{2-\lambda}{2}-i\frac{\omega}{2\kappa}\right)}e^{i\omega x}+
\frac{\Gamma\left(\frac{i\omega}{\kappa}\right)e^{-i\frac{\omega}{\kappa}\ln 2}}{\Gamma\left(\frac{\lambda+1}{2}+i\frac{\omega}{2\kappa}\right)\Gamma\left(\frac{2-\lambda}{2}+i\frac{\omega}{2\kappa}\right)}e^{-i\omega x}\right\}  ,\nonumber\\
\label{ueuo_asymptotic_1st_negativeX}
\eqn
for $x\to -\infty$, and 
\bqn
u_e(x) &\to& \Gamma\left(\frac12\right)\left\{\frac{\Gamma\left(\frac{-i\omega}{\kappa}\right)e^{i\frac{\omega}{\kappa}\ln 2}}{\Gamma\left(\frac{\lambda}{2}-i\frac{\omega}{2\kappa}\right)\Gamma\left(\frac{1-\lambda}{2}-i\frac{\omega}{2\kappa}\right)}e^{-i\omega x}+
\frac{\Gamma\left(\frac{i\omega}{\kappa}\right)e^{-i\frac{\omega}{\kappa}\ln 2}}{\Gamma\left(\frac{\lambda}{2}+i\frac{\omega}{2\kappa}\right)\Gamma\left(\frac{1-\lambda}{2}+i\frac{\omega}{2\kappa}\right)}e^{i\omega x}\right\} ,\nonumber\\
u_o(x) &\to& +\Gamma\left(\frac32\right)\left\{\frac{\Gamma\left(\frac{-i\omega}{\kappa}\right)e^{i\frac{\omega}{\kappa}\ln 2}}{\Gamma\left(\frac{\lambda+1}{2}-i\frac{\omega}{2\kappa}\right)\Gamma\left(\frac{2-\lambda}{2}-i\frac{\omega}{2\kappa}\right)}e^{-i\omega x}+
\frac{\Gamma\left(\frac{i\omega}{\kappa}\right)e^{-i\frac{\omega}{\kappa}\ln 2}}{\Gamma\left(\frac{\lambda+1}{2}+i\frac{\omega}{2\kappa}\right)\Gamma\left(\frac{2-\lambda}{2}+i\frac{\omega}{2\kappa}\right)}e^{i\omega x}\right\}  ,\nonumber\\
\label{ueuo_asymptotic_1st_positiveX}
\eqn
for $x\to +\infty$.

Based on Refs.~\cite{agr-qnm-Poschl-Teller-01,agr-qnm-Poschl-Teller-02}, one considers the bound state where $\omega$ is real.
We introduce the transformation
\bqn
\omega\to \omega'=\omega(\kappa')
\label{trans_PT_omega}
\eqn
together with those defined in Eq.~\eqref{trans_PT}, namely,
\bqn
\left\{\begin{array}{c}
x\to -ix'  \cr\\
\kappa \to i\kappa'  
\end{array}\right. . \nonumber
\eqn

After implementing the above substitution, $\kappa' x'$ remains real numbers, so the limits for the quantities such as $\kappa x$ and $z\equiv -\sinh^2\kappa x$ remain unchanged.
It is noted that the asymptotic forms for the wave functions transform from a bound state to out-going waves.
A tricky factor is that now $\lambda$ is complex owing to Eq.~\eqref{def_lambda}, which involves the substitution of $\kappa$.
Fortunately, one still has the asymptotic relation $\cosh^\lambda \kappa x(-z)^{-a}\to e^{i\frac{\omega}{\kappa}\ln 2}e^{-i\omega x}$, as it is easy to verify that the real part of $\lambda$ remains positive.

Subsequently, it is straightforward to find that
\bqn
A &=& {\Gamma\left(\frac{\lambda}{2}-i\frac{\omega}{2\kappa}\right)\Gamma\left(\frac{1-\lambda}{2}-i\frac{\omega}{2\kappa}\right)}, \nonumber \\
B &=& 2{\Gamma\left(\frac{\lambda+1}{2}-i\frac{\omega}{2\kappa}\right)\Gamma\left(\frac{2-\lambda}{2}-i\frac{\omega}{2\kappa}\right)} ,
\label{AB_ueuo}
\eqn
from which one also encounters the specific forms for $C, D$ given in Eq.~\eqref{f_mPT_CD} by comparing against Eqs.~\eqref{ueuo_asymptotic_1st_positiveX},
\bqn
C &=& \Gamma\left(\frac12\right)\left\{
\frac{\Gamma\left(\frac{\lambda}{2}-i\frac{\omega}{2\kappa}\right)\Gamma\left(\frac{1-\lambda}{2}-i\frac{\omega}{2\kappa}\right)}
{\Gamma\left(\frac{\lambda}{2}+i\frac{\omega}{2\kappa}\right)\Gamma\left(\frac{1-\lambda}{2}+i\frac{\omega}{2\kappa}\right)}
+
\frac{\Gamma\left(\frac{\lambda+1}{2}-i\frac{\omega}{2\kappa}\right)\Gamma\left(\frac{2-\lambda}{2}-i\frac{\omega}{2\kappa}\right)}
{\Gamma\left(\frac{\lambda+1}{2}+i\frac{\omega}{2\kappa}\right)\Gamma\left(\frac{2-\lambda}{2}+i\frac{\omega}{2\kappa}\right)}
\right\} \Gamma\left(\frac{i\omega}{\kappa}\right)e^{i\frac{\omega}{\kappa}\ln 2}, \nonumber \\
D &=& 2\Gamma\left(\frac12\right)\Gamma\left(\frac{-i\omega}{\kappa}\right)e^{i\frac{\omega}{\kappa}\ln 2} .
\label{CD_ueuo}
\eqn
We note that the above coefficients, $A$, $B$, $C$, and $D$ are not functions of the spatial coordinate $x$.
It is not difficult to find the residual of $C$ at the QNMs $\omega=\omega_n$ Eq.~\eqref{qnm_omega}
\bqn
\mathrm{Res}\left(C,\omega_n\right) =\left\{\begin{array}{cc}
\frac{(-1)^n 2i\kappa e^{i\frac{\omega_n}{\kappa}\ln 2}\Gamma\left(\frac12 \right)\Gamma\left(\frac{i\omega_n}{\kappa} \right)}{n!}
\frac{\Gamma\left(\frac{1-\lambda}{2}-i\frac{\omega_n}{2\kappa}\right)}{\Gamma\left(\frac{\lambda}{2}+i\frac{\omega_n}{2\kappa}\right)\Gamma\left(\frac{1-\lambda}{2}+i\frac{\omega_n}{2\kappa}\right)} & \ \ \mathrm{for}\ n=0,2,\cdots  \cr\\
\frac{(-1)^n 2i\kappa e^{i\frac{\omega_n}{\kappa}\ln 2}\Gamma\left(\frac12 \right)\Gamma\left(\frac{i\omega_n}{\kappa} \right)}{n!}
\frac{\Gamma\left(\frac{2-\lambda}{2}-i\frac{\omega_n}{2\kappa}\right)}{\Gamma\left(\frac{\lambda+1}{2}+i\frac{\omega_n}{2\kappa}\right)\Gamma\left(\frac{2-\lambda}{2}+i\frac{\omega_n}{2\kappa}\right)} & \ \ \mathrm{for}\ n=1,3,\cdots
\end{array}\right. .\label{ResComegan}
\eqn
As in Eq.~\eqref{qnm_PT}, we have only considered the modes whose real part is positive.

Following the discussions in Sec.~\ref{sec2.1}, one works out the expansions up to the next order in the limit $x\to +\infty$.
After some algebra, one finds
\bqn
u_e(x) &\to& \Gamma\left(\frac12\right)\left\{\left[1-\frac{\lambda(\lambda-1)}{\left(1+i\frac{\omega}{\kappa}\right)} e^{-2\kappa x}\right]\frac{\Gamma\left(\frac{-i\omega}{\kappa}\right)e^{i\frac{\omega}{\kappa}\ln 2}}{\Gamma\left(\frac{\lambda}{2}-i\frac{\omega}{2\kappa}\right)\Gamma\left(\frac{1-\lambda}{2}-i\frac{\omega}{2\kappa}\right)}e^{-i\omega x}\right. \nonumber\\
&+&\left.\left[1-\frac{\lambda(\lambda-1)}{\left(1-i\frac{\omega}{\kappa}\right)} e^{-2\kappa x}\right]\frac{\Gamma\left(\frac{i\omega}{\kappa}\right)e^{-i\frac{\omega}{\kappa}\ln 2}}{\Gamma\left(\frac{\lambda}{2}+i\frac{\omega}{2\kappa}\right)\Gamma\left(\frac{1-\lambda}{2}+i\frac{\omega}{2\kappa}\right)}e^{i\omega x}\right\} ,\nonumber\\
u_o(x) &\to& +\Gamma\left(\frac32\right)\left\{\left[1- \frac{\lambda(\lambda-1)}{\left(1+i\frac{\omega}{\kappa}\right)}e^{-2\kappa x}\right]\frac{\Gamma\left(\frac{-i\omega}{\kappa}\right)e^{i\frac{\omega}{\kappa}\ln 2}}{\Gamma\left(\frac{\lambda+1}{2}-i\frac{\omega}{2\kappa}\right)\Gamma\left(\frac{2-\lambda}{2}-i\frac{\omega}{2\kappa}\right)}e^{-i\omega x}\right.\nonumber\\
&+&\left.\left[1-\frac{\lambda(\lambda-1)}{\left(1-i\frac{\omega}{\kappa}\right)} e^{-2\kappa x}\right]\frac{\Gamma\left(\frac{i\omega}{\kappa}\right)e^{-i\frac{\omega}{\kappa}\ln 2}}{\Gamma\left(\frac{\lambda+1}{2}+i\frac{\omega}{2\kappa}\right)\Gamma\left(\frac{2-\lambda}{2}+i\frac{\omega}{2\kappa}\right)}e^{i\omega x}\right\}  .\nonumber\\
\label{ueuo_asymptotic_2nd_positiveX}
\eqn
Subsequently, in the place of Eq.~\eqref{f_mPT_CD}, we now have
\begin{equation}
\begin{aligned}
&\widetilde{C} \equiv C+\Delta C=C + \left(\Delta \widetilde{C}_1 + \Delta \widetilde{C}_2 \right)e^{-2\kappa x},  \\
&\widetilde{D} \equiv D+\Delta D=D + \Delta \widetilde{D} e^{-2\kappa x} ,
\end{aligned}\tag{\ref{CD_tilde_ueuo}}
\end{equation}
where
\bqn
{\Delta \widetilde{C}_1} &=& -\frac{\lambda(\lambda-1)}{\left(1-i\frac{\omega}{\kappa}\right)}
\frac{\Gamma\left(\frac{\lambda}{2}-i\frac{\omega}{2\kappa}\right)\Gamma\left(\frac{1-\lambda}{2}-i\frac{\omega}{2\kappa}\right)}
{\Gamma\left(\frac{\lambda}{2}+i\frac{\omega}{2\kappa}\right)\Gamma\left(\frac{1-\lambda}{2}+i\frac{\omega}{2\kappa}\right)} 
\Gamma\left(\frac12\right)\Gamma\left(\frac{i\omega}{\kappa}\right)e^{i\frac{\omega}{\kappa}\ln 2} ,\nonumber\\
{\Delta \widetilde{C}_2}&=& -\frac{\lambda(\lambda-1)}{\left(1-i\frac{\omega}{\kappa}\right)}
\frac{\Gamma\left(\frac{\lambda+1}{2}-i\frac{\omega}{2\kappa}\right)\Gamma\left(\frac{2-\lambda}{2}-i\frac{\omega}{2\kappa}\right)}
{\Gamma\left(\frac{\lambda+1}{2}+i\frac{\omega}{2\kappa}\right)\Gamma\left(\frac{2-\lambda}{2}+i\frac{\omega}{2\kappa}\right)}
\Gamma\left(\frac12\right)\Gamma\left(\frac{i\omega}{\kappa}\right)e^{i\frac{\omega}{\kappa}\ln 2} , \nonumber \\
{\Delta \widetilde{D}} &=& -\frac{2\lambda(\lambda-1)}{\left(1+i\frac{\omega}{\kappa}\right)}
\Gamma\left(\frac12\right)\Gamma\left(\frac{-i\omega}{\kappa}\right)e^{i\frac{\omega}{\kappa}\ln 2} .
\label{CD_tilde_delta_ueuo}
\eqn

\section{Expansion coefficients for the waveforms in the disjointed square barrier potential}\label{appB}

The calculations for a single squire barrier are straightforward, and the reflection and transmission amplitudes are found to be~\cite{agr-qnm-16, agr-qnm-instability-18}
\bqn
C &=& e^{-i\omega d_1}\frac{iV_1\sin\left(\sqrt{\omega^2-V_1}d_1\right)}{2\omega \sqrt{\omega^2-V_1}},\nb \\
D &=& e^{i\omega d_1}\frac{2\omega \sqrt{\omega^2-V_1}\cos\left(\sqrt{\omega^2-V_1}d_1\right)-i\left(2\omega^2-V_1\right)\sin\left(\sqrt{\omega^2-V_1}d_1\right)}{2\omega\sqrt{\omega^2-V_1}}. \label{CDstep}
\eqn
Besides, we also note that a comprehensive study of QNMs as poles of the transition matrix can be found in~\cite{qm-scattering-05}.

To derive $E$ and $F$, one first makes use of the unitary of the transit matrix~\cite{book-blackhole-Frolov}, which implies the following relations between the reflection and transmission amplitudes when the frequency takes real values
\bqn
\widetilde{E} &=& D(d_1\to d_2, V_1\to V_2) ,\nb \\
\widetilde{F} &=& -C(d_1\to d_2, V_1\to V_2)^* , \label{EFtilde_step}
\eqn
then displace the potential barrier by $x_\mathrm{step}$ alone the $x$ axis:
\bqn
E &=& \widetilde{E}  ,\nb \\
F &=& e^{2i\omega x_\mathrm{step}}\widetilde{F}  , \label{EFdisplacement_step}
\eqn
and we find
\bqn
E &=& e^{i\omega d_2}\frac{2\omega \sqrt{\omega^2-V_2}\cos\left(\sqrt{\omega^2-V_2}d_2\right)-i\left(2\omega^2-V_2\right)\sin\left(\sqrt{\omega^2-V_2}d_2\right)}{2\omega\sqrt{\omega^2-V_2}}  ,\nb \\
F &=& e^{2i\omega x_\mathrm{step}}e^{i\omega d_2}\frac{iV_2\sin\left(\sqrt{\omega^2-V_2}d_2\right)}{2\omega \sqrt{\omega^2-V_2}}  . \label{EF_step}
\eqn

\bibliographystyle{h-physrev}
\bibliography{references_qian}

\end{document}